\documentclass{sn-jnl}

\usepackage{amsmath}
\usepackage{amssymb}
\usepackage{booktabs}
\usepackage{pythonhighlight} 
\usepackage[utf8]{inputenc}
\usepackage{natbib} 
\usepackage{appendix}
\usepackage{threeparttable}  
\usepackage{multirow}

\begin{document}

\title{Across Time and (Product) Space: A Capability-Centric Model of Relatedness and Economic Complexity}

\author[1]{\fnm{Ziang} \sur{Huang}}\email{zh435@cam.ac.uk}
\author*[2]{\fnm{Huashan} \sur{Chen}}\email{chenhs@cass.org.cn}

\affil[1]{\orgname{University of Cambridge}. \it Cambridge, UK\normalfont}

\affil[2]{\orgname{Chinese Academy of Social Sciences}. \it Beijing, China\normalfont}


\maketitle
\begin{abstract}
EEconomic complexity - a group of dimensionality-reduction methods that apply network science to trade data - represented a paradigm shift in development economics towards materializing the once-intangible concept of capabilities. Measures such as the Economic Complexity Index (ECI) and the Product Space have proven themselves as robust estimators of an economy's subsequent growth; less obvious, however, is why they are so. Despite ECI drawing its micro-foundations from a combinatorial model of capabilities, where a set of homogeneous capabilities combine to form products and the economies which produce them, such a model is consistent with neither the fact that distinct product classes draw on distinct capabilities, nor the interrelations between different products in the Product Space which so much of economic complexity is based upon. \\

In this paper, we extend the combinatorial model of economic complexity through two innovations: an underlying network which governs the relatedness between capabilities, and a production function which trades the original binary specialization function for a product-level output function. Using country-product trade data across 216 countries, 5000 products and two decades, we show that this model is able to replicate both the characteristic topology of the Product Space and the complexity distribution of countries' export baskets. In particular, the model transforms measures of economic complexity into direct measures of the capabilities held by an economy, both improving the informativeness of the ECI in predicting economic growth and enabling an interpretation of economic complexity as a proxy for productive structure via CES-like factor substitutability.
\end{abstract}

\footnote{As we were preparing this manuscript, we became aware of concurrent work by Hidalgo and Stojkoski, which was uploaded to arXiv on July 24th, 2025. Their paper independently explored a very similar idea to ours: a model of multiple capabilities, coupled with an output function rather than a binary specialization function, which provides theoretical grounding for both the Product Space and economic complexity. \\ \\
Excepting a single section in Section 7 (Discussion and conclusion) which explicitly compares the two papers, this paper was not produced with knowledge of their work, and not been modified in light of it; additionally, we hope that the simultaneity in our research efforts underscores the importance of the central theme of both papers - that of the role of capabilities in economic complexity. For a more detailed comparison between the two papers, see Section 7.}

\section{Introduction}
The advent of \it{economic complexity }\normalfont - a wide-ranging, interdisciplinary field related principally to development economics, encompassing a family of data-driven methods rooted in network science - stands today as perhaps the most promising culmination of two fundamental trends in development economics from the 1950s to the present. The first, a gradual realization that the role of endogenous knowledge, or \it capabilities, \normalfont played a more significant and more directly observable role than the first models of economic growth - which treated it less as genuine economic variable and more as divinely bestowed manna - had suggested (Solow 1956; Swan 1956; Lucas 1988; Romer 1989); and the second, a decisive break from one-size-fits-all, coarse-grained growth models of the past and towards fine-grained models that treated countries, products, and even individual firms as distinct and distinctly modellable. \\

It is at this intersection where the central thesis of economic complexity emerged: that previously-theoretical constructs of product varieties and the knowledge needed to produce them are empirically observable through data. As previously seen with models such as the Grossman-Helpman quality ladder model (Grossman and Helpman 1991), the notion that products are differentiable in their quality had already been codified into endogenous growth theory; however, recognition that \it specific \normalfont product varieties extant in real-world data are themselves differentiable in requisite knowledge originates from a series of seminal papers primarily by Hidalgo and Hausmann between 2000-2010. Motivated by the notion of \it{cost discovery} \normalfont (Hausmann and Rodrik 2002) - an extension of positive R\&D spillovers in Romer's original endogenous growth model to the discovery of new varieties of goods - it is first assumed that different goods require different productivity thresholds for entrepreneurs to undertake their discovery, then that such a threshold can be quantitatively proxied by the average income level (termed $PRODY$) of all countries who specialize in exporting that product (Hausmann, Hwang and Rodrik 2005). The idea of using country-product exports to proxy for the productivity requirements of producing a product, by then known as \it{complexity}\normalfont, was further expanded into a recursive formulation - the Method of Reflections - in which a product's complexity was the average complexity of the countries specializing in it, and a country's complexity was the average complexity of the products it specialized in (Hidalgo and Hausmann 2009). \\

Simultaneous to the advent of economic complexity was the advent of the \it{principle of relatedness} \normalfont - the notion that products could be described by the similarity between the capabilities required to produce them, proxied by a single empirically observed statistic: how frequently two products are co-specialized in by the same economy. Together, complexity and relatedness formed the backbone of a framework of development whose granularity reached the level of individual - and, more importantly, specific and tangible - products, whose capability requirements were encapsulated in economic complexity and whose relationships to one another were encapsulated in relatedness and the network it brought forth: the Product Space. \\

The field of development economics has made great use of economic complexity methods. The Economic Complexity Index (ECI) was found to be a more robust predictor of a country's economic growth rate versus traditional metrics such as GDP per capita or investment-to-GDP ratios over a 20-year period (Hidalgo and Hausmann 2009); similarly, the concepts of product relatedness and the Product Space were shown to predict future diversification prospects for a country's export baskets on the level of individual products (Hidalgo et al. 2007), exemplifying the notion of \it{path dependence} \normalfont - that a country's future prosperity depends heavily on previous specializations - central to endogenous growth theory. Precisely due to the fact that economic complexity was an empirically-grounded, rather than theoretically-grounded, methodology that could be applied to any economy and product, economic complexity methods have been fruitfully applied to pinpointing products exacerbating income inequality (Hartmann et al. 2017; Chu and Hoang 2020), to identifying the carbon footprint of products and supporting green development paths (Fraccascia et al. 2018; Neagu and Teodoru 2019), and to economies at any scale, from country-level studies to studies of specific provinces, sub-national regions, and even towns (Mealy et al. 2019). Iterative improvements to the Method of Reflections, including the non-linear Fitness Algorithm (Tacchella et al. 2012) and simpler formulations such as a combinatorially-derived complexity measure based on capabilities (Inoua 2023) have outperformed the original ECI on forecasting future growth; further studies of export diversification paths have expanded the principle of relatedness to acknowledge commonalities between products such as shared labor inputs (Schetter et al. 2024), related technologies via patents (Balland et al. 2022), and downstream customer linkages (Bahar et al. 2017), and uncovered a wealth of evidence both in the realm of path-dependent (relatedness-driven) and path-defying (relatedness-resisting) behavior (Neffke et al. 2011; Coniglio et al. 2021). \\

Throughout its plethora of applications over the years, the crux of economic complexity remains exactly the same today as it was nearly two decades ago - to observe what was once unobservable: the capabilities required for countries to produce certain products. Indeed, Hausmann's original formalization of economic complexity (Hausmann and Hidalgo 2011) conceived a simple combinatorial model where products and countries were both represented as subsets of a finite string of homogeneous capabilities. Even the simplest variant of such an "ingredients-in-a-recipe" model - where capabilities are homogeneous, and countries produce a good if and only if it has accrued all of its requisite capabilities - can be analytically solved to replicate the most important stylized facts characterizing the country-product export network, including the fact that low-complexity products are exported by most countries and high-complexity products are exported by only a few countries (Hausmann and Hidalgo 2011). With a few slight modifications to the model, such as introducing substitutability between capabilities (Lei and Zhang 2014) or positing that economies will abandon redundant low-complexity products as it develops (van Dam and Frenken 2022), further empirical facts like "the hump" - an inverse U-shaped relationship between country income and export diversification (Cadot et al. 2011) - also begin to emerge. \\

Therefore, the question becomes: are methods of economic complexity a true measure of economic capabilities and aggregate knowledge, the missing piece to endogenous growth theory and the Solow-Swan model before it? Despite attempts to extend economic complexity beyond trade data to separate indices of technology (Stojkoski et al. 2023) and research (Balland et al. 2022), two roadblocks remain in bridging the gap between the ECI as it stands towards a truly comprehensive measure of capabilities. \\

The first is a reconciliation of quality and variety. Though studies have shown that product varieties classified under similar product classes (e.g. machinery, chemicals, metals) often have similar complexities (Felipe et al. 2012), and that complexity methods can be applied directly to more "genotypic" input-output data, such as data showing the mix of occupations necessary for the production of each product (Schetter et al. 2024), the fact that complexity places every product variety on a single scale leads to a loss of information on how products relate to one another - in essence, the information encoded by the Product Space. Is a more complex product truly a product requiring more capabilities, and are products with similar complexities similar in the capabilities they use? Neither question has seen a solution which is either empirically or theoretically clear. \\

The second is clarity on the nature of capabilities themselves. Though the metaphor of "capabilities combining to create products" has been nearly universally borrowed as a micro-foundation for economic complexity, it falls short in genuinely connecting the mathematical underpinnings of economic complexity to the picture of capability combinations which such a micro-foundation paints. For a sub-field of development economics whose primary objective is to transform what was previously intangible - aggregate knowledge and capabilities - into something tangible for every product and every economy, economic complexity deserves to be provided a more transparent model of capabilities that simultaneously captures both its keystones: the principle of relatedness and the notion of complexity. \\

In this paper, we propose a modification of the original combinatorial model underlying economic complexity as derived by Hidalgo and Hausmann by relaxing two of the original assumptions - first, that capabilities are functionally identical and unsubstitutable; and second, that countries can only specialize in a product if it holds all its requisite capabilities, and that specialization exists in only one of two binary states, "specialized" or "not specialized". The key innovation of our model is the introduction of an underlying Capability Space, a symmetric block-form matrix which quantifies the relatedness between pairs of capabilities, with more related capabilities having a higher likelihood of combining to form the same product. \\

Section 2 will briefly introduce and summarize the methods of economic complexity, including revealed comparative advantage, relatedness, the Method of Reflections; it will then present a mathematical interpretation that links product complexity and the Product Space; Section 3 will briefly describe our dataset and methodology for processing data. Section 4 will introduce our model; Section 5 will characterize the topology of the Product Space and its complex-network properties, and provide an application of our model to the reproduction of the Product Space; and Section 6 will apply the model to countries' export baskets and the ECI proper, demonstrating first that the model is able to explain the distribution of products a country exports, and subsequently that the best-fitting parameters to the model for a country can serve as both a powerful indicator of the substitutability and returns to scale of the country's capabilities, and a means of transforming economic complexity into measures of directly observed capabilities which are more statistically informative than the ECI in predicting economic growth. Finally, Section 7 discusses the position of this work in the related literature, particularly given a concurrent contribution by Hidalgo and Stojkoski, and concludes the paper.

\section{Methods}

Studies in economic complexity until now have primarily utilized one of two foundational methods originally presented by Hidalgo and Hausmann: the Product Space and relatedness, and measures of economic complexity such as the ECI and Product Complexity Index (PCI). This section will summarize the methodology underlying both the Product Space and economic complexity, then discuss mathematical links between them and suggest reasons on why they should be viewed as a unified framework rather than two disparate theories.

\subsection{Exports and export specialization}
The Product Space and most measures of economic complexity share a common emphasis on country-product export data. In particular, both methods employ Revealed Comparative Advantage (RCA) as a proxy for specialization (Balassa 1965), defined as follows for countries denoted $c$, products denoted with $p$, and exports of country $c$ of product $p$ denoted $x_{cp}$:

\begin{equation}
   \text{RCA}_{c,p} = \frac{x_{cp}/\sum_{p}x_{cp}}{\sum_{c}x_{cp}/\sum_{c}\sum_p x_{cp}} = \frac{\text{share of $p$ in total exports of $c$}}{\text{share of $p$ in total world exports}}
\end{equation}

In particular, the threshold for a country \it specializing \normalfont in producing $p$ is $RCA_{c,p}> 1$, where the share of $p$ in $c$'s total exports exceeds the global average.

\subsection{Ubiquity and diversity}

Immediately following from the above, we note that information on country-product specialization can be conveniently encoded in a \it country-product matrix \normalfont $M$:

\begin{equation}
    M_{cp} = \begin{cases}
        1, \text{RCA}_{c,p} > 1 \\
        0\ \text{otherwise}
    \end{cases}
\end{equation}

where $c$ and $p$ range from $1$ to the total number of countries and products respectively. This leads to two more metrics which can be thought of as proto-measures of complexity for countries and products respectively. The first is \it diversity\normalfont, defined and denoted for a country $c$ as $k_{c,0}$:

\begin{equation}
    k_{c,0}=\sum_{p} M_{cp} = \text{number of products country $c$ has RCA in}
\end{equation}

and the second, \it ubiquity\normalfont, defined and denoted analogously for a product $p$ as $k_{p,0}$:

\begin{equation}
    k_{p,0}=\sum_{c} M_{cp} = \text{number of countries having RCA in product $p$}.
\end{equation}

\subsection{The Product Space and relatedness}

The Product Space is a network representing the interconnections between exported product varieties representing the \it proximit y\normalfont between all pairs of products. For any two products $i$ and $j$, define the \it proximity \normalfont between $i$ and $j$ - denoted $\phi_{ij}$ - as follows, symmetrized by a minimum:

\begin{equation}
\phi_{ij} = \text{min}\{P(\text{RCA}_{c,i}|\text{RCA}_{c,j}), P(\text{RCA}_{c,j} | \text{RCA}_{c,i})\}
\end{equation}

or, in other words, the conditional probability that country $c$ has RCA in $i$ given it has RCA in $j$, or vice versa, whichever is smaller; here the dummy variable of country $c$ is to be interpreted as calculating this conditional probability across all countries, i.e.

\begin{equation}
  P(\text{RCA}_{c,i} | \text{RCA}_{c,j}) = \frac{\text{number of times $i$ and $j$ are co-specialized in by the same country}}{\text{number of times any country specializes in $j$}}.
\end{equation}
Alternatively, we may formulate the above through the country-product matrix $M$ as well as diversity $k_{c,0}$ and ubiquity $k_{p,0}$:

\begin{equation}
(M^T M)_{ij} = \sum_{c} M_{ci} M_{cj} 
\end{equation}
where
\begin{equation}
M_{ic} M_{cj} = \begin{cases}
    1,\ \text{$c$ specializes in both $i$ and $j$} \\
    0\text{ otherwise}
\end{cases}
\end{equation}
and as such
\begin{equation}
    \begin{aligned}
        \frac{(M^T M)_{ij}}{k_{j,0}} &= \frac{\text{number of times $i$ and $j$ are co-specialized in by the same country}}{\text{number of times any country specializes in $j$}} \\
        &= P(\text{RCA}_{c,i} | \text{RCA}_{c,j})
    \end{aligned}
\end{equation}
and, denoting the adjacency matrix of the Product Space network as $\Phi$, we have
\begin{equation}
    \phi_{ij} = \min\{\frac{(M^T M)_{ij}}{k_{j,0}}, \frac{(M^T M)_{ij}}{k_{i,0}}\}
\end{equation}
and thus 
\begin{equation}
    \Phi = \min\{ U^{-1}M^T M, (U^{-1}  M^T M)^T\}
\end{equation}
where the $\text{min}$ operation is taken element-wise on the two matrices of identical dimension ($p \times p$), and $U$ is the diagonal ubiquity matrix whose non-diagonal elements are zero and whose diagonal elements $U_{pp}$ are the ubiquities of products $p$. \\ 

Of particular interest in forecasting future export diversification prospects is a country-level metric called \it density\normalfont, defined as the average relatedness of the current export specializations $c$ to a new product $i$ and denoted $\omega_{ci}$:
\begin{equation}
    \omega_{ci} = \frac{\sum_{p} M_{cp}\phi_{pi}}{k_{c,0}} = (D^{-1}M\Phi)_{ci}
\end{equation}
where $D$ is the diagonal diversity matrix whose non-diagonal elements are zero and whose diagonal elements $D_{cc}$ are the diversities of countries $c$.

\subsection{Measures of economic complexity}

At its core, the ECI aims to compress the dimensionality of diversity-ubiquity data by combining information on the diversity of a country's exports with the ubiquity of its exported products. This is encapsulated by the following pair of iterative equations (Hidalgo and Hausmann 2009), termed \it the Method of Reflections\normalfont, where $k_{c,N}$ and $k_{p,N}$ denote the ECI and PCI for country $c$ and product $p$ after the $N$th iteration, $M$ remains the country-product matrix, and $k_{c,0}$ and $k_{p,0}$ diversity and ubiquity for country $c$ and product $p$ respectively:

\begin{equation}
    \begin{cases}
        k_{c,N} = \frac{1}{k_{c,0}}\sum_{p}M_{cp}k_{p,N-1} \\ 
        k_{p,N} = \frac{1}{k_{p,0}}\sum_{c}M_{cp}k_{c,N-1}.
    \end{cases}
\end{equation}

It has been observed that if we denote 
\begin{equation}
    \tilde{M} = D^{-1}MU^{-1}M^T
\end{equation}
and 
\begin{equation}
    \hat{M} = U^{-1}M^TD^{-1}M
\end{equation}
for the matrices underlying the iterative calculations of ECI and PCI respectively, then we simply obtain:
\begin{equation}
    \begin{cases}
        \vec{k}_{c,2N} = \tilde{M}^N \vec{k}_{c,0} \\
        \vec{k}_{p,2N} = \hat{M}^N \vec{k}_{p,0} 
    \end{cases}
\end{equation}
The steady state of this iterative approach is given by the vectors which solve the following equations:
\begin{equation}
    \begin{cases}
        k^*_{c} = \tilde{M} k^*_{c} \\ 
        k^*_{p} = \hat{M} k^*_{p}
    \end{cases}
\end{equation}
or, in other words, the eigenvectors of $\tilde{M}$ and $\hat{M}$ respectively. Note that, as a consequence of the row-stochasticity of the matrices involved and the Perron-Frobenius theorem (Mealy et al. 2019), the ones vector is a steady-state solution to the above equations; as it would be completely uninformative to assign all countries the same ECI through the ones vector, we take the second-largest eigenvector of both $\tilde{M}$ and $\hat{M}$ as the ECI and PCI vectors respectively, usually normalized.

\subsection{Linking relatedness and complexity}

While the economic interpretation of economic complexity remains comparatively elusive, more light has been shed on the mathematical meaning of the Method of Reflections and its implications (Kemp-Benedict 2014; Mealy et al. 2019). \\ 

The first point of interest is the mathematical justification for why economic complexity exists as a distinct measure from diversity and ubiquity. In their original paper, Hidalgo and Hausmann note that "successive generations of... measures of economic complexity ($k_{c,N}$)... are increasingly good predictors of growth", suggesting that the Method of Reflections enables a synthesis of diversity and ubiquity which captures information that diversity alone cannot; indeed, as shown in a subsequent study (Kemp-Benedict 2014), the eigenvector representing ECI is \it mathematically orthogonal \normalfont to diversity. \\ 

The second point of interest is the precise mathematical meaning of the eigenvectors which represent ECI and PCI: most significantly, the interpretation of the ECI vector as a method of spectral clustering which provides an approximate solution to the problem of partitioning the country-product specialization graph into two balanced components (Mealy et al. 2019). The ECI vector is \it the \normalfont unique vector which best assigns each country a numerical score, such that countries which are more strongly connected in the country-country specialization similarity graph, with adjacency matrix given by $MU^{-1}M^T$, are assigned more similar scores; and along a similar vein, the PCI vector assigns products more strongly connected in the graph with adjacency matrix $\hat{M} = U^{-1}M^TD^{-1}M$ similar scores. \\

We clarify this conceptual link by directly comparing the matrix underlying the Product Space with the matrix used to calculate PCI. Recall that the adjacency matrix of the Product Space was defined in matrix form as 
\begin{equation}
\Phi = \min\{ U^{-1}M^T M, (U^{-1}  M^T M)^T\}
\end{equation}
with entries $\phi_{ij}$ understood as the conditional probability that a country specializes in $j$ given that it specializes in $i$ (or vice versa, whichever is smaller). We note that the matrix $\hat{M} = U^{-1}M^TD^{-1}M$, used as the adjacency matrix for the product-product similarity graph in calculating PCI, differs from $\Phi$ only by the normalization matrix $D^{-1}$. It follows that the Product Space arises from a special case of the derivation of the PCI eigenvector, in which $D$ is a multiple of the identity matrix 
\begin{equation}
    D = \begin{bmatrix}
        d & 0 & ... & 0 \\
        0 & d & ... & 0 \\
        \vdots & \vdots & \ddots & \vdots \\
        0 & 0 & ... & d
    \end{bmatrix} = dI
\end{equation}
with $\hat{M} = U^{-1}M^T D^{-1}M = \frac{1}{d} \Phi$ reducing to a constant multiple of the Product Space. In such a theoretical scenario - where countries all specialize in the same number of products - the PCI eigenvector would also represent an approximately optimal spectral clustering of the Product Space; products which have strong links in the Product Space would be characterized by similar PCIs. This gives us a  framework for understanding why relatedness via the Product Space and economic complexity are inextricably connected:
\begin{itemize}
    \item The (unsymmetrized) Product Space, represented by the adjacency matrix $\Phi = U^{-1}M^T M$, measures product co-export conditional probabilities without regard for the countries which specialize in them.
    \item The diversity-normalized matrix $U^{-1}M^T D^{-1}M$, whose second largest eigenvector represents PCI, measures the exact same notion of product co-export conditional probabilities, but under normalization by country diversity: co-specialization by a very specialized, least-diversified economy is considered more significant because it would indicate that both products arise from the same set of (relatively limited) capabilities. 
    \item The PCI vector is mathematically equivalent to an approximately optimal spectral clustering of this diversity-normalized matrix, which generalizes the Product Space and reduces to it (in its unsymmetrized form) when all countries are equally diverse. In such a scenario, the PCI vector represents an approximately optimal partitioning of the Product Space as well. 
\end{itemize}

\section{Data and methodology}
The following sections will make use of the newest version of the BACI dataset of international trade flows (Gaulier and Zignago 2010), comprised of bilateral trade flows disaggregated at the exporter-importer-product level across more than 200 countries and 5008 products according to the HS92 Harmonized System 6-digit classification of goods. All explanatory variables in regressions are sourced from the World Bank's World Development Indicators (World Bank 2025) and the Global Macro Database (Muller et al. 2025), which contains data from 1994 to 2023. All results in Section 5 and 6 of the paper use data from four years, spread evenly within the two decades from 2000 to 2020: 2000, 2005, 2010 and 2015. We avoid any data from 2020 onwards due to disruptive effects from the COVID-19 pandemic. \\ 

We \textbf{do not} follow the standard practice of excluding all countries with total exports not exceeding \$1 billion USD, and all products with total exports not exceeding \$100 million USD. Removing low-flow countries and products would remove some amount of noise at the cost of creating both flickering of countries and products in and out of the study, as a product excluded in one year could be included in another year, as well as distorting the underlying structure of the Product Space, specifically the degree, weight and centrality distributions. Additionally, zero trade flows are important to a later part of the paper (Section 6, which involves using the production function of the model to model export baskets); excluding them would artificially alter the distribution of country exports. 

\section{A model of related capabilities}

Let us begin with a recollection of the Hidalgo-Hausmann combinatorial model of capabilities, based on the following assumptions:
\begin{itemize}
    \item There exist a finite set of homogeneous capabilities of size $N_a$, denoted $A$; no two capabilities are substitutable.
    \item All products and countries can be represented as subsets of $A$, or alternatively, as columns of binary matrices $\mathbf{C}_{ca}$ and $\mathbf{P}_{pa}$ respectively, whose entries are one if the country $c$ (product $p$) possesses (requires) a capability $a$ and zero otherwise.
    \item The probability that a country or product possesses or requires a capability $a$ is independent of that of other capabilities.
    \item A country can only produce a product if it possesses all of its requisite capabilities. If a country possesses every requisite capability of a product, it will produce that product; in other words, if 
    \begin{equation}
        \sum_{a} \mathbf{C}_{ca} \mathbf{P}_{pa} = \sum_{a} P_{pa}
    \end{equation}
    denoted by the \it Leontief operator\normalfont
    \begin{equation}
        \mathbf{C}_{ca} \odot \mathbf{P}_{pa} = 1
    \end{equation}
    then country $c$ will produce product $p$.
\end{itemize}

Mean-field estimations using this simple model proved remarkably capable in replicating several stylized facts significant to development economics, the relationship between diversity and ubiquity prime among them; where this model is found lacking is its inability to capture the relationships between products. This fact becomes even more glaring given the fundamental ties between the Product Space and economic complexity. As Hidalgo and Hausmann themselves point out in their original paper, "products require the combination of several inputs... a shoe manufacturer and a circuit board company both need accountants and a cleaning crew... the circuit board manufacturing plant, on the other hand... requires people skilled in photo-engraving... which have no use in the shoe factory." It is therefore very difficult to justify a model where capabilities are regarded as completely homogeneous and unrelated, and where the possibility of similar products requiring similar capabilities is entirely dismissed. \\

Thus, the core modification to the model we propose is a relaxation of the assumption of homogeneous, unrelated capabilities via an underlying \it Capability Space\normalfont. Suppose that there exist a total of $N_a$ capabilities, $N_c$ countries and $N_p$ products; define the Capability Space, denoted $\Phi^C$, as the $N_a\times N_a$ matrix whose entries $\Phi^C_{ij}$ indicate a measure of \it relatedness \normalfont between two capabilities $i$ and $j$, and whose diagonal $\Phi^C_{ii}$ is comprised of purely ones. Formally, this relatedness is defined as 
\begin{equation}
    \Phi^C_{ij} = P(i \in P | j \in P)
\end{equation}
given an arbitrary product $P$ which is only known to contain $j$ and has $|P| \geq 2$. It is important to keep in mind throughout the following section that capabilities and products in this model do \textbf not \normalfont correspond exactly to real-world capabilities and products, nor do we intend them to; the model aims to capture the mechanisms driving the structure and topology of the Product Space, rather than specific products within it. The goal of the next two sections will be to show that, given well-calibrated starting parameters, this relatively simple model of a latent Capability Space can be used to generate a set of synthetic products that accurately reflects the macroscopic structure of the real-world Product Space. \\ 

Under the paradigm of the Capability Space, we reconsider three aspects of the original model: the procedure under which combinations of capabilities form products, the definition of economic complexity and product proximity within this capabilities-based model, and the expression of the production function for a country $c$ and a product $p$ given $\mathbf{C}_{ca}$ and $\mathbf{P}_{pa}$.  

\subsection{Capability blocks}
A simple assumption for the structure of the Capability Space is that its adjacency matrix $\Phi^C$ takes a block-matrix form, in the spirit of a stochastic block model (Holland et al. 1983). Specifically, we have
\begin{equation}
    \Phi^C = \begin{bmatrix}
    \Phi^C_{1,1} & \Phi^C_{1, 2} & ... & \Phi^C_{1, n} \\
    \Phi^C_{2,1} & \Phi^C_{2,2} & ... & \Phi^C_{2,n} \\
    \vdots & \vdots & \ddots & \vdots \\
    \Phi^C_{n,1} & \Phi^C_{n,2} & ... & \Phi^C_{n,n}
\end{bmatrix}
\end{equation}

with blocks $1, 2, ..., n$, where each $\Phi_{i,j}$ is shorthand for a matrix of the form

\begin{equation}
  \Phi_{i,j} = \phi^C_{i,j} \times \begin{bmatrix}
    1 & 1 & ... & 1 \\
    1 & 1 & ... & 1 \\
    \vdots & \vdots & \ddots & \vdots \\
    1 & 1 & ... & 1 
\end{bmatrix}.  
\end{equation}

for some scalar constant $\phi^C_{i,j}$. \\
 
We make this assumption for two reasons. The first is the nature of products in the Product Space. Though the precise topology of the Product Space has not been studied at length, it is well-understood that the Product Space possesses both a core-periphery as well as an evident community structure (Hidalgo et al. 2007) in which goods such as textiles form tightly-knit clusters; it is then reasonable to infer a Capability Space in which certain groups of capabilities are associated with certain classes of products (e.g. looms for textiles), and are thus more heavily linked to one another than to other classes of capabilities. The second is purely practical; a block-matrix Capability Space means that the model will involve far fewer parameters, and when the Capability Space is a single block, the model reduces to the original Hidalgo-Hausmann combinatorial model. \\

We now turn to calibrating the block-matrix structure from empirical Product Space data. Recall from Section 2 that the Product Complexity Index is \bf the \normalfont vector representing the approximately optimal spectral partition for a network capturing product-to-product similarities into distinct spectral clusters; indeed, under unique conditions, this network is exactly equal to the Product Space. \newpage

\begin{center}
    \begin{figure}[h]
        \includegraphics[width=13cm]{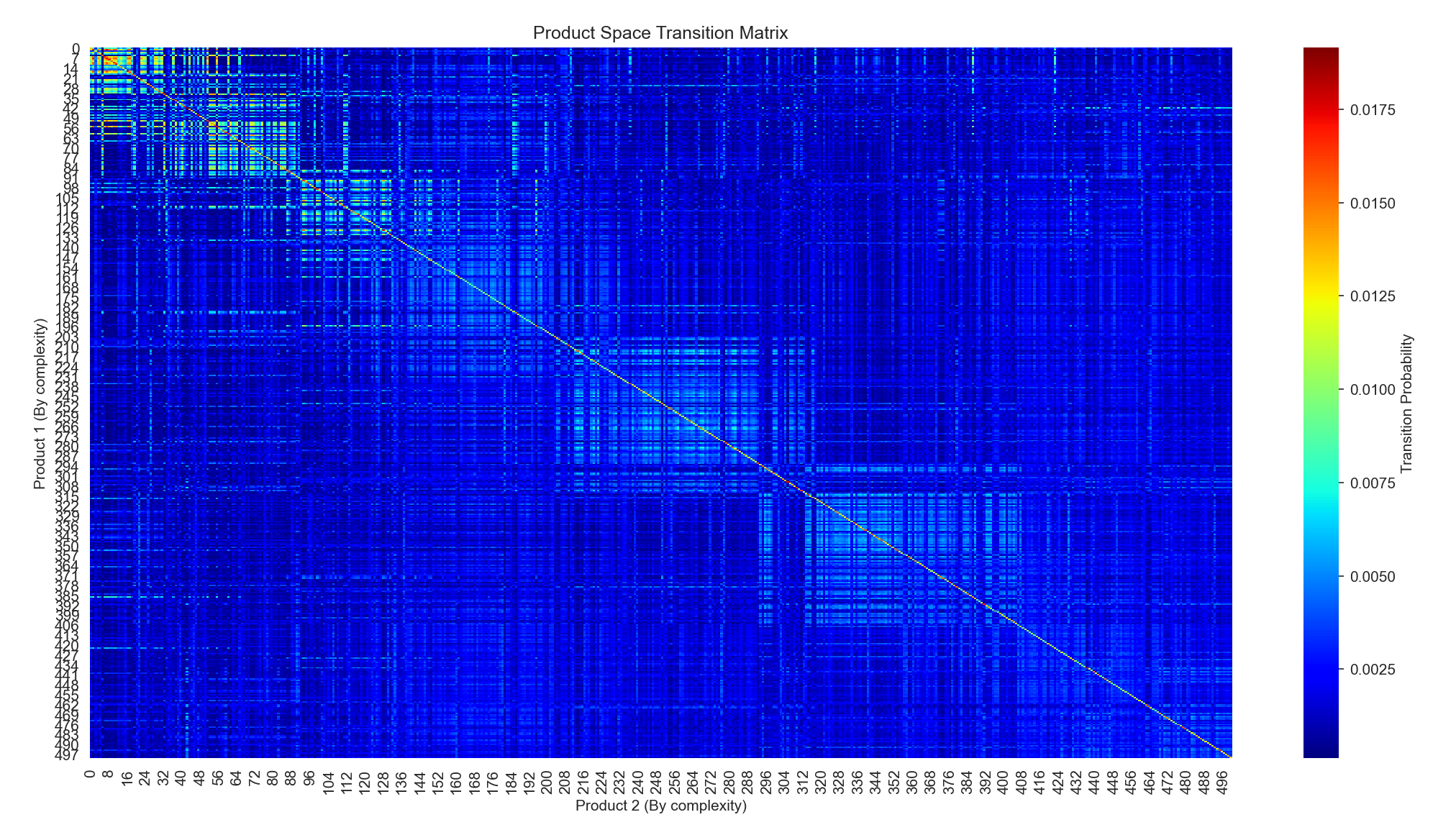}
        \caption{Heatmap of the Product Space (row-normalized) using trade data for 5008 products according to the HS92 6-digit classification in 2005. Products are ordered by complexity on both axes. Brighter colors indicate more intense proximities; note the clear presence of block-like structures which indicate close connections between products with similar complexities}
    \end{figure}
\end{center}

As such, we fit a Gaussian mixture model (GMM) with $n$ components to the PCI data in a specific year (e.g. 2005), with $n$ a parameter of the model, to identify clusters within the product space (which we assume to roughly correspond with capability clusters); we interpret each fitted component $N_i = N(\mu_i, \sigma_i^2)$ as a block in the Capability Space, such that in total we have $n$ blocks within $\Phi^C$. \\

To further simplify the model assumptions, we assume that every block $N_i$ falls into one of two types, determined by whether $\mu_i$ is less than mean PCI $\mu$: periphery ($\mu_i < \mu$), and core ($\mu_i > \mu$). Blocks are identical to other blocks of the same type in both the number of capabilities within each block (denoted $N^p_{a}$ and $N^c_{a}$ for low and high complexity blocks respectively), the value of each within-block element ($\phi^p_{w}$ and $\phi^c_{w}$), and the value of each between-block element ($\phi^p_{b}$ and $\phi^c_{b}$); we assume that $\phi^p_{b} \leq \phi^p_{w} \leq 1$ and $\phi^c_{b} \leq \phi^c_{w} \leq 1$. \\

Generating a product $p$ into the model entails randomly selecting a block from the $n$ blocks present in $\Phi^C$, each with probability equal to $\lambda_i$, or their weight in the fitted GMM; a starting capability for $p$ is chosen at random from that block, and the total number of capabilities $K_{p,0}$ required to produce $p$ assumed to follow the Gaussian distribution $K_{p,0} \sim N_i = N(\mu_i, \sigma_i^2)$ above. Capabilities are then chosen iteratively from the set of capabilities not yet in $p$ until there are a total of $K_{p,0}$ capabilities in $p$, according to a process described in the next section. As we define one proxy for the complexity of a product to be the number of capabilities required to produce it, this conveniently allows for product complexity in our model to be drawn from the same distribution as the best-fitting GMM to the empirical PCI distribution:

\begin{equation}
    E(K_{p,0}) = \sum_{i=1}^n P(\text{$p$ is in block $i$}) N(\mu_i, \sigma_i^2) = \sum_{i=1}^n \lambda_i N_i \approx E(PCI)
\end{equation}

(It is worth noting that GMMs can be understood as a form of "soft clustering" of PCI values, similar to how PCI itself is a spectral clustering of the product-to-product similarity network; due to this, GMMs of $2$ or more components are a consistently good fit for the PCI distribution spanning two decades for 5008 products from 2000-2023. GMM blocks should \textbf not \normalfont be interpreted as corresponding to real-world capabilities, but as a means of calibrating the boundary conditions of the model to align with empirical reality. Precise details on this are found within Appendix A.1.)

\subsection{Product formation}
A product $p$ is represented formally as a non-empty subset of the set of (distinct) capabilities $A$; this subset is understood as the set of capabilities required to produce $p$. Given a known value for its size, denoted $K_{p,0}$, the set of capabilities comprising $p$ is generated iteratively as follows:
\begin{enumerate}
    \item Suppose that at the first step, $p$ contains a single (randomly assigned) capability $a_0$. 
    \item At step $i = 2, 3, ..., K_{p,0}$, preferentially attach new capabilities to $p$ via the following rule: the probability of capability $a* \notin P$ being attached to $p$ at this step is
    \begin{equation}
        P(a^* \text{ is chosen next}) = \sum_{a\in p}\frac{1}{i} \Phi^C_{aa^*} 
    \end{equation}
    equalling the average proximity of the current capability basket of $p$ to capability $a^*$ in the Capability Space. The probabilities across all unchosen capabilities are normalized to sum to $1$, and a new capability is randomly chosen.
    \item After step $K_{p,0}$, stop. 
\end{enumerate}
The rationale for this iterative probabilistic approach is as follows. First, it reinforces the principle that products which fall into the same product class (such as textile products) share similar capabilities. If two products start with capabilities belonging to the same block, then their next capability would also be likely to originate from that block, creating a preferential-attachment dynamic; this models the existence of products which lie firmly within a single sector. Second, it introduces the possibility of cross-sector products which employ capabilities from multiple blocks in the Capability Space; in particular, consider the following scenario 
\begin{equation}
    p=\{a_0\}
\end{equation}
in the starting step, where $a_0$ is a capability belonging to the textile block. Suppose that $\Phi^C$ takes a form such that it is far more probable for the next capability attached to $p$, denoted $a_1$, to also originate from the textile block; however, if instead $a_1$ was chosen from the block of capabilities representing the chemicals sector, then in the next step with $p=\{a_0, a_1\}$, the textile block would no longer be preferred for the next capability and the product would be more likely to require capabilities from a wider variety of blocks. Intuitively speaking, this indicates that the more capabilities a product is known to require, the more likely it will branch out into other blocks outside its starting block by sheer chance, or even exhaust all capabilities from its starting block; and thus that products requiring more capabilities will require capabilities from a more diverse range of sectors. 

\subsection{Defining complexity and proximity}
Like the original model, we model countries $c$ as subsets of the set of capabilities $A$; these subsets are understood to be the capabilities that a country possesses. As such, we define the following: 

\subsubsection{Complexity}
In the original model, the \it complexity \normalfont of a set of capabilities $A^* \subset A$, representing either a product or country, was calculated simply as the number of capabilities in $A^*$, denoted $K_{A^*,0}$: $K_{A^*,0} = |A^*|$, reflecting the intuition that countries (products) which possess (require) more capabilities were more complex. This is essentially analogous to the \it zeroth-order \normalfont measures of economic complexity in the Method of Reflections: both diversity $k_{c,0}$ and ubiquity $k_{p,0}$ are transformed from counting the number of specializations to counting the number of capabilities. In subsequent sections, we will employ both this measure of complexity as well as the following measure of \it first-order complexity\normalfont, in the spirit of the original Method of Reflections; denoting the complexity of a specific capability $a$ by $K_{a,1}$ and the first-order complexity of $A^*$ by $K_{A^*,1}$, we have:
\begin{equation}
    \begin{cases}
        K_{a,1} = \frac{1}{|P_{a\in p}|} \sum_{p \in P_{a\in p}} K_{p,0} \\
        K_{A^*, 1} = \frac{1}{K_{A^*, 0}} \sum_{a \in A^*} K_{a,1}
    \end{cases}
\end{equation}
where the average complexity of a capability $a$ is defined as the average complexity of all products which use it (denoted by the set $P_{a \in p}$), and the first-order complexity of a product or country $A^*$ as the average complexity of all capabilities it requires. \\ 

As Section 2 states, the vector of country diversities is mathematically orthogonal to the eigenvector representing ECI; as such, a first-order complexity measure which synthesizes the number of capabilities (diversity of countries) with the complexity of these capabilities (ubiquity of products) is necessary to capture further information present in the data. We therefore opt to use first-order complexity in all remaining sections of this paper. \\ 

\subsubsection{Proximity}
For any two subsets $A_1, A_2 \subset A$, where $A_1, A_2$ can represent either countries or products, define the \it proximity \normalfont between $A_1$ and $A_2$ to be the average pairwise proximity (inherently symmetric as defined below) between the capabilities contained within $A_1$ and $A_2$:

\begin{equation}
    \phi(A_1, A_2) = \frac{1}{K_{A_1,0} K_{A_2,0}}\sum_{a_1 \in A_1} \sum_{a_2 \in A_2} \phi^C_{a_1 a_2} 
\end{equation}

\subsection{The production function}
Suppose that a product $p$ requires a capability $p_i$. Instead of a binary Leontief production function, we suppose that production in $p$ is increasing in a country $c$'s endowment in $p_i$: the more that $c$ is endowed in capabilities related to $p_i$, or $p_i$ itself, the more it is able to easily make use of the skills, technology, knowledge, etc. represented through the capability $p_i$, and thus produce $p$. As such we define
\begin{equation}
    \phi(c, \{p_i\}) = \frac{1}{K_{c,0}}\sum_{a \in c} \Phi^C_{ap_i}
\end{equation}

as a measure of the average proximity of $c$ to capability $p_i$, roughly analogous to the concept of density in the Product Space. One natural production function for the quantity of $p$ produced by $c$ is

\begin{equation}
    Q(c,p) = \frac{\alpha}{K_{c,0}}\sum_{i=1}^{K_{p,0}} \phi(c, \{p_i\})
\end{equation}

averaging the proximity of $c$ to all capabilities required by $p$, with $\alpha$ a country-specific constant representing total factor productivity. Closer inspection reveals that this is a special case (where the elasticity of substitution $\sigma$ is infinite) of the following CES production function:

\begin{equation}
Q(c,p) = \alpha(\sum_{i=1}^{K_{p,0}} \frac{1}{K_{p,0}} \phi(c,\{p_i\})^\rho)^\frac{\nu}{\rho}
\end{equation}

where $\nu$ is the returns to scale parameter, and $\rho$ a parameter determining substitutability between capabilities such that the elasticity of substitution $\sigma$ equals $\frac{1}{1-\rho}$. Calculating
\begin{equation}
   R(c,p) = \frac{Q(c,p)}{\sum_{p\in P}Q(c,p)},
\end{equation}
or the share of $p$ in the production of $c$, allows us to directly estimate the export distribution of countries on a product-by-product level. Note that $\alpha$ cancels out and is thus not a parameter in the model. \\

This generalized production function embeds within itself other production functions that have previously been used to model capabilities; in particular, the Leontief-like production function of the original Hidalgo-Hausmann model is recoverable from this generalized form when $\rho$ approaches $-\infty$, $\nu$ is one and $\sigma$ approaches zero (unsubstitutable capabilities). For a full derivation, see Appendix A.2.

\subsection{Concluding the model}

In this section, we have presented a combinatorial model of capabilities that extends the original Hidalgo-Hausmann model in two ways - by introducing heterogeneous, interrelated capabilities through the existence of an underlying Capability Space, whose block-matrix structure is inferred from the PCI data via a Gaussian mixture model by exploiting the mathematical link between PCI and spectral clustering; and by introducing a production function founded upon the CES production function which allows for variable endowment in capabilities, of which the original model's Leontief production function is a special case. \\ 

The following sections will apply this model to three empirical benchmarks foundational to the field of economic complexity. First, we will show that the model can recover the empirically observed topology of the Product Space purely from PCI data, including the edge weight (proximity) and centrality distributions as well as the community structure, in four years spread across two decades (2000, 2005, 2010, and 2015). Second, we will show that this model of capabilities partially explains the variation in countries' export distributions when $\rho$ and $\nu$ are allowed to vary; the values of $\rho$ and $\nu$ themselves, particularly their relationship with ECI, are themselves meaningful as indicators for the productive structure of an economy, allowing an interpretion of ECI as a measure of the productive structure and thus explaining the mechanism behind why ECI predicts growth. Finally, using our zeroth-order and first-order measures of complexity, we demonstrate that the set of capabilities a country possesses, inferred from a best-fit to its export data, creates a more informative measure of complexity that better predicts economic growth compared to ECI.

\section{Results: Modeling the Product Space}

Though more general surveys of the Product Space have noted its core-periphery and community structure (Hidalgo et al. 2007), we find that the Product Space is characterized by three unusual network properties that distinguish it from other complex networks: a left-skewed, unimodal degree and centrality distribution where most nodes (products) in the network are connected to a majority of other nodes (products), a right-skewed weight distribution where most pairs of products are only weakly connected (exhibit low proximity), and the fact that its structure is closely intertwined with economic complexity, with nodes representing more complex products being more central to the network and nodes with similar complexity connecting more strongly to one another. In the following section, we will examine each property in turn and show that these properties arise from our model of capabilities presented above; though the analysis below uses data from 2005, we will also present results for data in 2000, 2010 and 2015 in Appendix B. \newpage

\begin{center}
    \begin{figure}[h]
        \includegraphics[width=14cm]{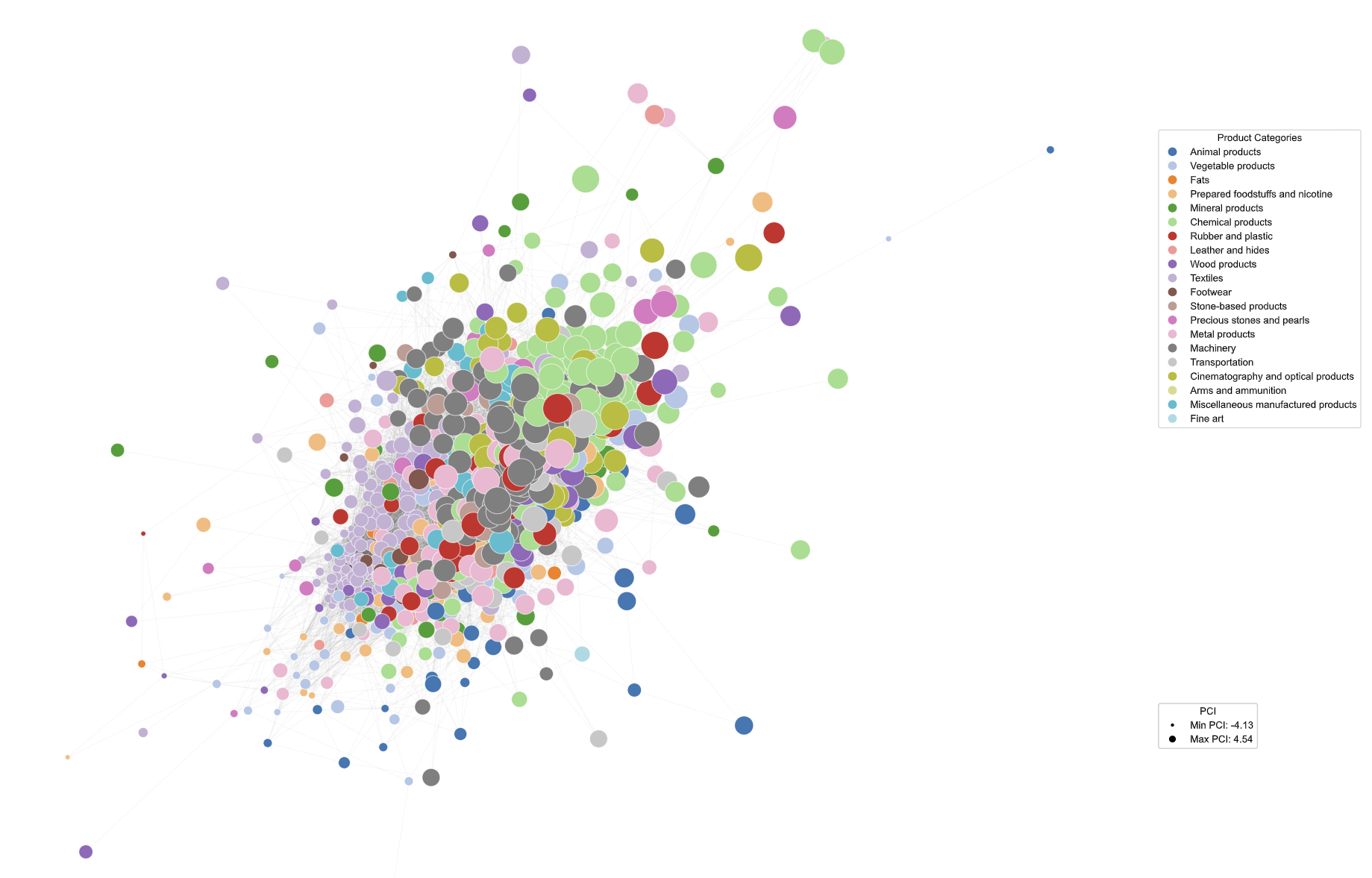}
    \caption{The Product Space, visualized from data in 2005. To prevent visual clutter, only 1000 out of 5008 nodes are shown (randomly chosen); larger nodes have higher product complexity, and nodes are color-coded by HS92 classification (under one of 21 broad chapters). Note that high-complexity products, usually in sectors like chemicals, machinery and metals, occupy a visible core, while low-complexity products like textiles occupy a periphery}
    \end{figure}
\end{center}

\subsection{Topology of the Product Space}

Here we present calculated values for the following standard network summary statistics and centrality measures: density, global clustering coefficient, modularity (Newman 2006), and eigenvector centrality, and further characterize the distributions of edge weight (proximity), node centrality, and node degrees. \\

These statistics are calculated for the Product Space in 2005, with results for additional years (2000, 2010, 2015) available in Appendix B.1 and showing roughly stable values for all measures below. Note that edges with zero weight are treated as completely absent; this is to prevent the degree distribution from becoming constant, and the weight distribution from being swamped by zero weights. \\

As shown, the Product Space is a very strongly connected network, with more than 90\% of all pairs of nodes being connected (density $> 0.9$); this is also reflected by the extremely high clustering coefficient across a 20-year period. The Product Space is weakly modular, with a maximum modularity of $0.12$ when partitioned via the Leiden algorithm. It is further characterized by a right-skewed edge weight distribution, with a unimodal peak at roughly 0.1 to 0.2, and a left-skewed eigenvector centrality, degree centrality, and strength centrality distribution, with most nodes being central (in particular, the degree of a majority of nodes is very close to the maximum possible degree of 5008). We note that according to a two-sample KS test at the 5\% significance level, no parametric distribution is a good fit for this degree distribution; for more details, see Appendix B.1.

\begin{table}[htbp]
\centering
\caption{Summary statistics for the Product Space (2005)}

\begin{tabular}{lcccc}

& \textbf{Number of nodes} & \textbf{Density} & \textbf{Clustering coefficient}  & \textbf{Modularity}\\
\midrule
\textbf{2005}                                                    & 5008 & 0.9044     &   0.9296     &  0.111  \\ 
\end{tabular}
\end{table}

\begin{table}[h]
\centering
\caption{Distribution statistics for the Product Space (2005)}

\begin{tabular}{lcccccc}

Statistic & \textbf{Mean} & \textbf{Median} & \textbf{Mode} & \textbf{IQR} & \textbf{Skew} & \textbf{Kurtosis} \\
\midrule
\textbf{Centrality} & 0.589 & 0.610 & 0.003 & 0.225 & -0.619 & 0.245 \\
\textbf{Weight} & 0.154 & 0.139 & 0.167 & 0.125 & 0.947 & 0.989 \\ 
\textbf{Degree} & 4528 & 4702 & 4767 & 4822 & -3.49 & 16.5 \\ 
\end{tabular}
\end{table}

In total, the summary statistics for the distributions point to a series of characteristics entirely distinct from most complex networks. Contrary to the scale-free properties of networks such as the World Trade Network (Serrano and Boguna 2003) in which degree distributions follow a power law of the form $k^{-\omega}$ with $\omega \approx 1.6$, underlaid by models such as the Barabasi-Albert model of preferential attachment (Barabasi and Albert 1999), it is clear that the degree distribution of the Product Space is extremely left-skewed and unimodal, while the weight distribution is also significantly right-skewed and unimodal - not distributed according to any power law. This points to a fundamental difference in the mechanism through which the Product Space attains its topology. \newpage

\begin{figure}[h]
    \centering
    \includegraphics[width=12cm, height=8cm]{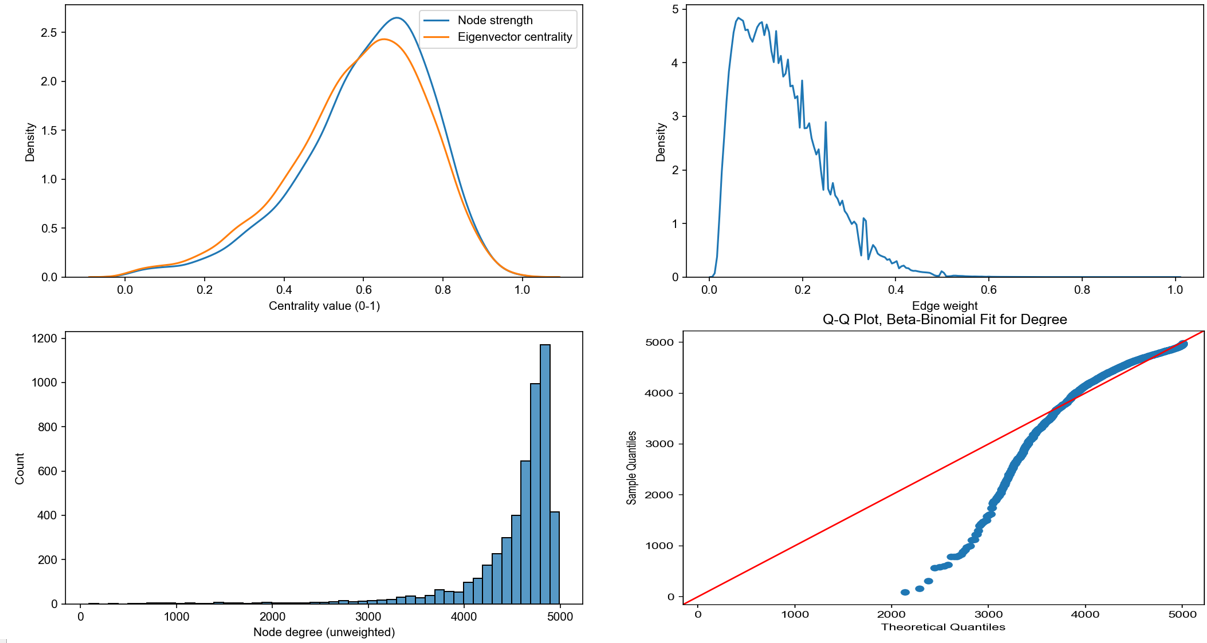}
    \caption{Product Space degree and centrality distributions. Top left: centrality distributions. Top right: proximity distributions. Bottom left and right: node degree distributions are well-approximated by a binomial distribution at the right end, but poorly at the left end}
\end{figure}

\subsection{Product complexity in the Product Space}

We examine the role of product complexity in shaping the structure of the Product Space through three lens: centrality (nodes representing more complex products are more central, i.e. have higher eigenvector centrality), assortativity (nodes with similar PCI are more strongly connected), and community structure (naive binning by dividing products into equally-sized PCI bins attains a modularity similar to the Leiden algorithm). \\ 

For centrality, we directly calculate the $R$-value (the correlation coefficient and not the $R$-squared, because relationships may be negative) between product PCI and eigenvector centrality, obtaining a significant positive value; for assortativity, we calculate the $R$ between the difference in PCI between two products and their proximity $\Phi_{ij}$, obtaining a significant negative value (the more different the PCI, the weaker the proximity); and for community structure, we classify products into communities into $n$ bins which divide the PCI scale (roughly $-4.5 < PCI < 4.5$) evenly into $n$ segments, choosing the maximum value of modularity attained using different values of $n$ for $1 \leq n \leq 20$. We also include the modularity attained through simply classifying products into communities based on their HS92 chapter code (see Product Space figure above) as a reference point, and show that this method of naively binning by PCI outpeforms the HS92 classification and only slightly underperforms the approximate maximum modularity attainable via the Leiden algorithm when considering the top $X\%$ of edge weights (where $X$ is $100, 90, 80, ..., 10, 1$.)
\begin{table}[h]
\centering
\caption{Product complexity in the Product Space (2005)}
\begin{tabular}{lcccc}

    & \textbf{PCI-centrality correlation} & \textbf{$\Delta$PCI-proximity correlation} & \textbf{PCI modularity} \\
\midrule
\textbf{2005} &  0.344   &   -0.362     &  0.089  \\ 
\end{tabular}
\end{table}

\begin{figure}[h]
    \centering
    \includegraphics[width=8cm, height=6cm]{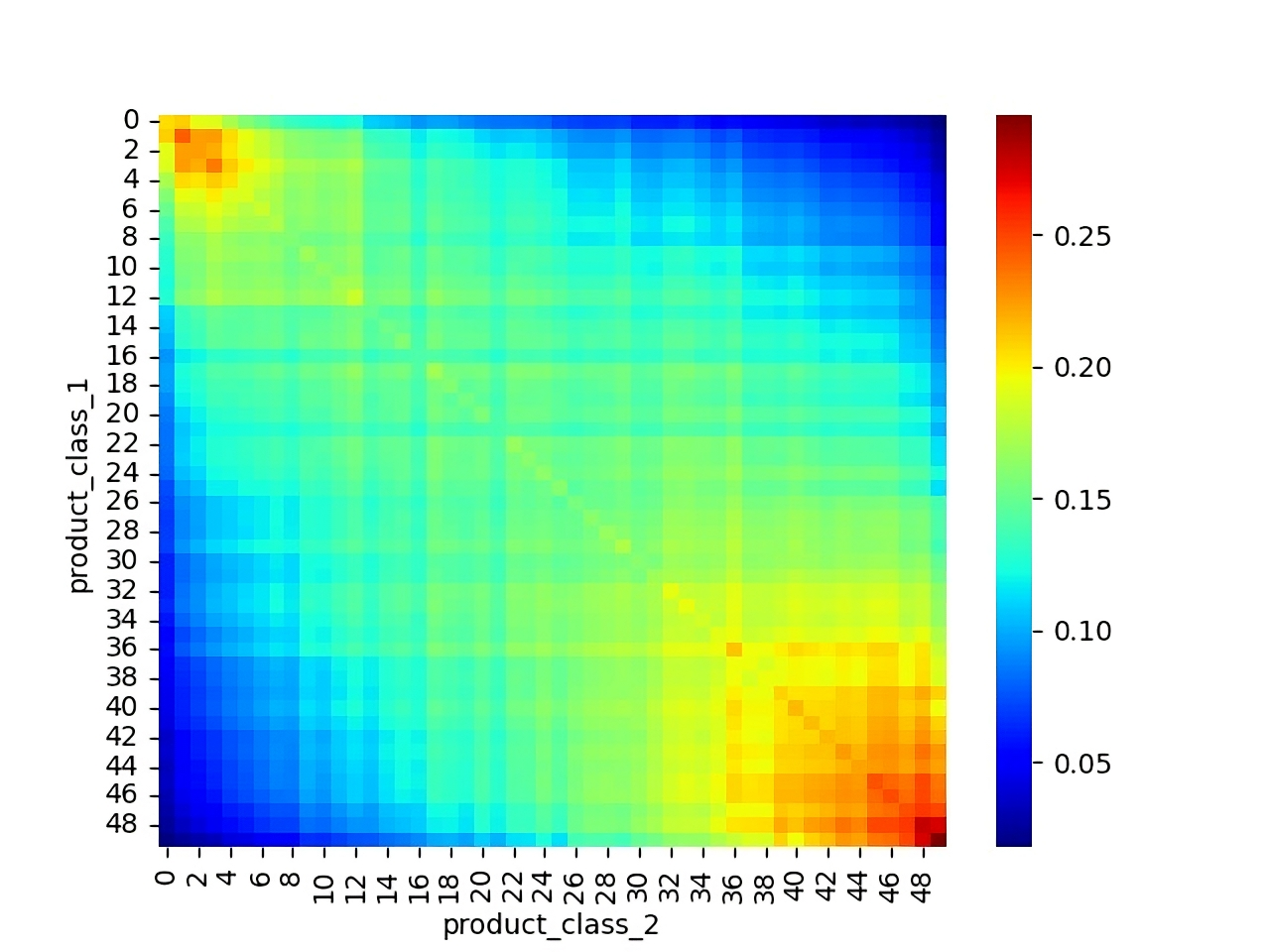}
    \caption{Heatmap of the Product Space, classified into 50 evenly-sized PCI bins; average pairwise proximity between products in each bin is shown in the heatmap, with entries closer to the diagonal (similar PCIs) having higher intensities}
\end{figure}

\subsection{Results for the model}

The following subsection will present results obtained from a numerical simulation of the model introduced in Section 3. We tune parameters minimally via an evolutionary algorithm, CMA-ES (Hansen 2016), to minimize Kolmogorov-Smirnov distance between the generated and empirical \textbf{weight distribution only}, with population size $\lambda = 20$ for $50$ generations. Additionally, as stated in Section 4, all zero-weight edges are eliminated from the network to heterogenize node degree; by model construction, a zero-weight edge would mean an average capability proximity of zero and thus be highly unlikely. As such, to ensure a meaningful comparison between the degree distribution of the two networks, we find the percentile that the first non-zero edge represents in the list of edge weights sorted in ascending order for the empirical Product Space; we then eliminate all edges in the generated Product Space below that percentile. For a full table of the tuned parameters and their ranges, see Appendix B.3. \\

It is worth addressing the potentially circular reasoning inherent in trying to reconstruct properties of the Product Space, including its relationship with the PCI, when our model already uses a GMM fitted on the empirical PCI distribution as a boundary condition. Though the use of the GMM does guarantee that the distribution of the number of capabilities of products matches the PCI distribution, it does \textbf not \normalfont guarantee any other property; \textit{a priori}, we do not assume any relationship between the complexity of a product and its well-connectedness in the Product Space (degree), nor its relationship with other similarly-complex products (PCI assortativity), nor can the weight or centrality distribution be trivially deduced from the PCI distribution. Therefore, the goal of this simulation is essentially to infer both the globally emergent and the local properties of the network listed above (centrality, degree, modularity etc.) from two of its lowest-level structural properties: the weight distribution and the distribution of product PCIs, equivalent to the distribution of the number of capabilities in our model. \\

Results for the Product Space in 2005 using a constant distribution for $\Phi^C$ are as follows; for precise values of parameters found by the optimization algorithm, as well as sensitivity analyses involving the model parameters (results across different years, variations on GMM components, introduction of random noise and capability heterogeneity), see Appendix B.3. Remarkably, the simplest variant of the model (no heterogeneity between capabilities in the same block) can lead to a near-complete replication of emergent higher-level topological properties of the Product Space via only its weight and PCI distribution, excepting the centrality distribution, where its left-skewed tendency emerges from the model but not its precise characteristics. This is an empirical limitation of the model that can potentially be addressed by increasing the granularity specified; the current results are nowhere near optimal and serve only as proof-of-concept. 

\begin{table}[h]
\centering
\caption{Simulation results for the Product Space (2005)}
\begin{tabular}{lcc}
\textbf{Property name} & \textbf{Empirical value}  & \textbf{Simulated value} \\
\midrule
Modularity (Leiden) & 0.111 & 0.125 \\
PCI modularity & 0.089 & 0.068 \\
$\Delta$PCI-proximity correlation & -0.362 & -0.412 \\
PCI-centrality correlation & 0.344 & 0.249 \\
Degree distribution D-statistic (KS Test) &n/a & 0.055* \\
Weight distribution D-statistic (KS Test) &n/a& 0.060* \\
Centrality distribution D-statistic (KS Test) &n/a & 0.307 
\end{tabular}
\begin{tablenotes}
\small
\item (*The null hypothesis that the simulated and empirical distributions are identical cannot be rejected at the 5\% significance level (D $\textless$ 0.061).) \\
\end{tablenotes}
\end{table}
\newpage
\begin{figure}[h]
    \centering
    \includegraphics[width=14cm, height=7cm]{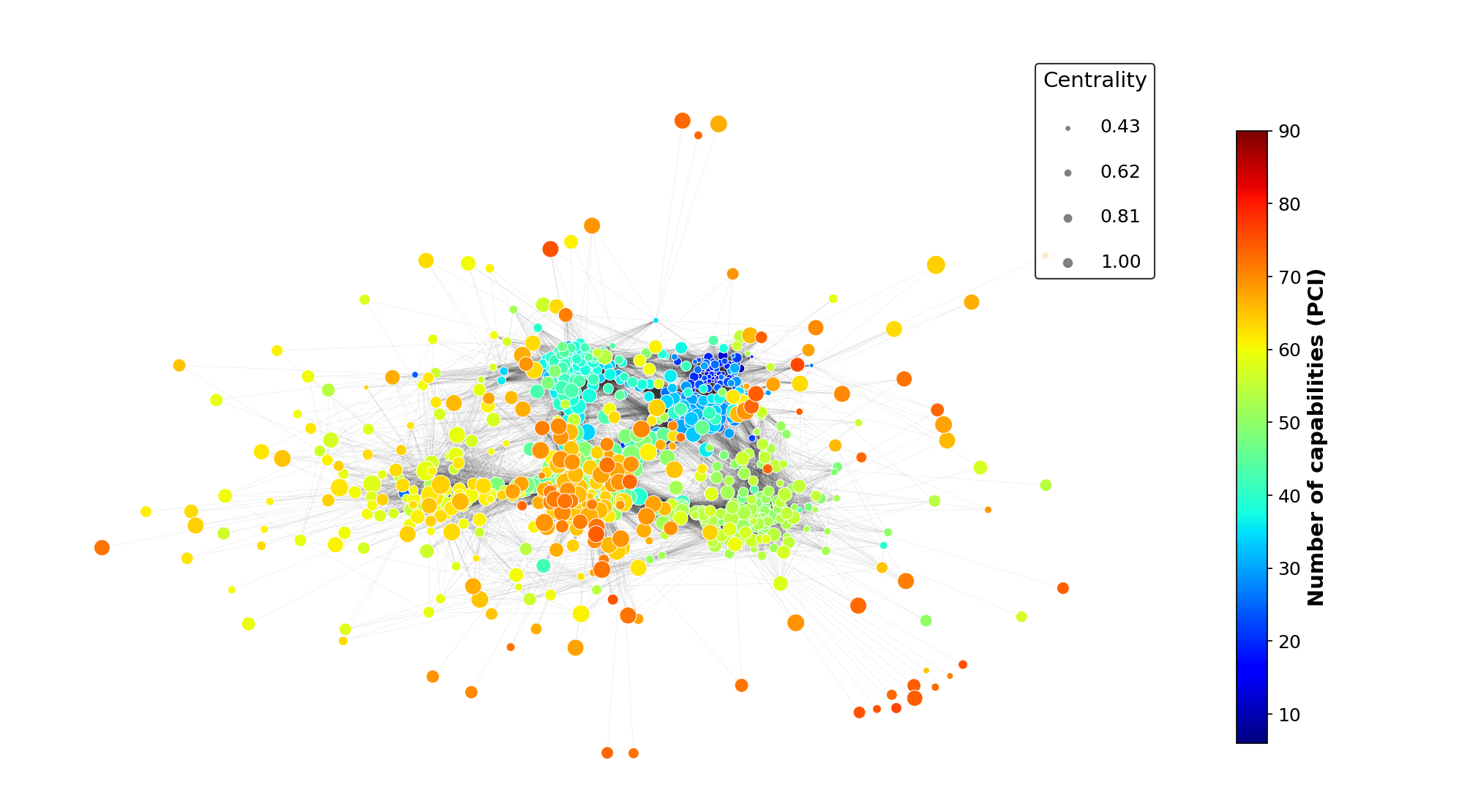}
    \caption{Visualization of the simulated product space using the parameters above. Only the top 10\% of edges by proximity are kept to avoid visual clutter. Nodes correspond to products; node colors correspond to different values of PCI (number of capabilities required); node sizes correspond to different values of centrality. Note the evident core-periphery structure, as well as how similarly-colored (similar-PCI) nodes cluster together}
\end{figure}

The fact that the proximity between periphery blocks is far higher - in fact, nine times higher - than that between core blocks is indicative of an intriguing asymmetry between "periphery" and "core", or less-complex and more-complex, capabilities; this may suggest a model of capabilities where capabilities often used in more complex products (core capabilities) are specialized and do not combine easily with other specialized capabilities, while capabilities used in less complex products (periphery capabilities) are general and can combine with other general capabilities. While we are cautious of drawing conclusions about the nature of real-world capabilities based on our model, the structure and characteristics of the Capability Space remains an interesting prospect for future study. \newpage

\begin{figure}[h]
    \centering
    \includegraphics[width=12cm, height=6cm]{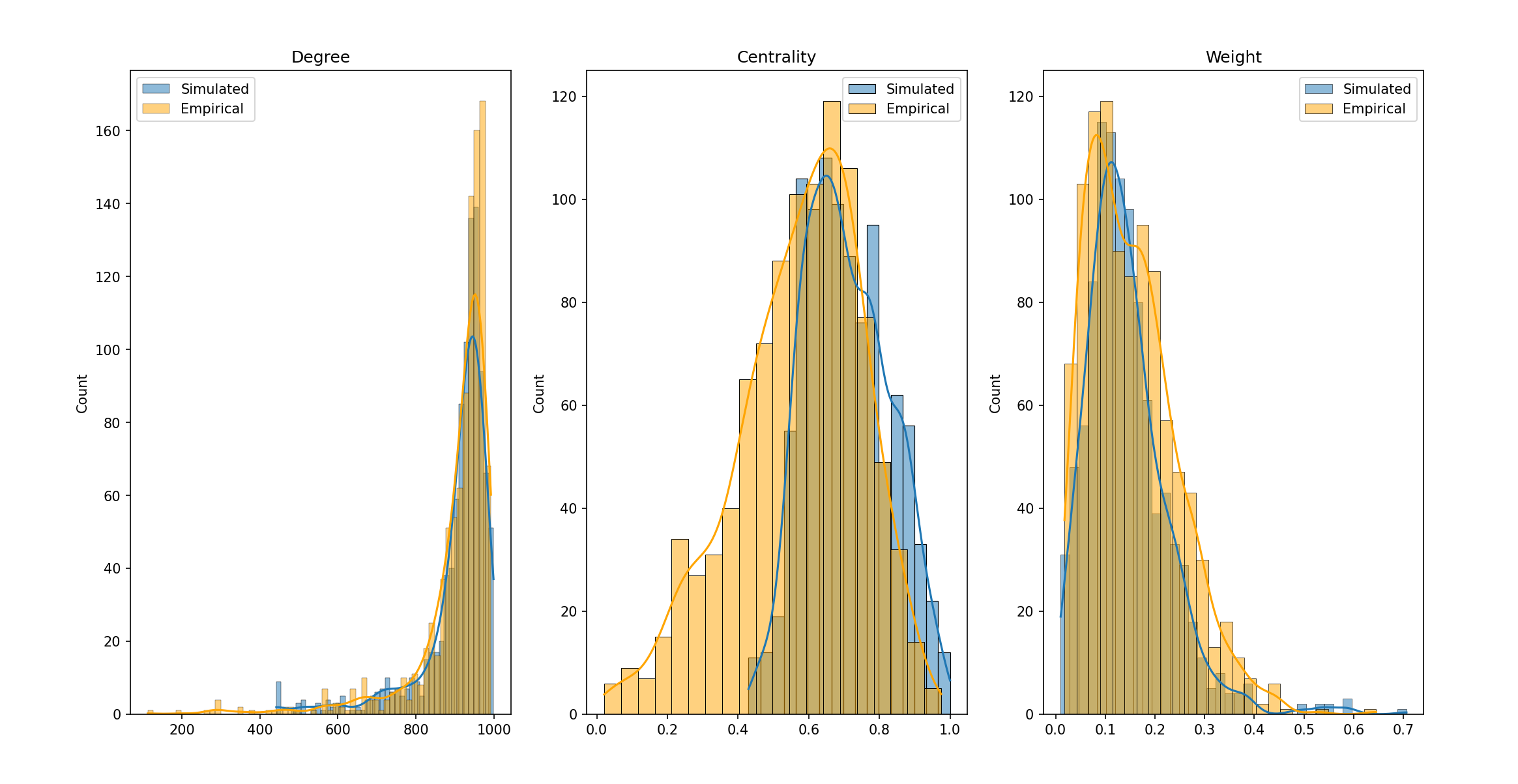}
    \caption{Simulated versus empirical distributions for the Product Space in 2005. Note goodness-of-fit for degree and edge weight distributions, but less so for centrality; however, the model does capture the left-skewed nature of the centrality distribution and its correlation with PCI}
\end{figure}

\section{Results: Modeling economic complexity}

We now turn to the key result: interpreting the ECI in terms of the latent capabilities and productive structure of an economy. The Capability Space calibrated in Section 5 describes how capabilities combine to form products. Given this, we can now invert the relationship: for any country with observed export shares across products, we infer which capabilities it likely possesses. Specfically, we apply the model's CES production function to empirically observed country export baskets in 2005, isolate a smoothed PCI-export distribution (export share vs. PCI of product), and infer countries' latent capability baskets by minimizing a cost function - the KL divergence between the predicted and actual export-PCI distribution - while also allowing the elasticity of substitution $\sigma$ and the returns to scale parameter $\nu$ to vary. We will then demonstrate that the values of $\rho$ and $\nu$ which provide this best-fit correlate robustly with the economic development of the country in question - $\rho$ positively, and $\nu$ negatively. \\

Finally, we show that this underlying set of latent capabilities inferred for each country yields two measures of complexity (number of capabilities and average capability complexity), introduced as zeroth-order and first-order measures of complexity in Section 3 respectively, which are highly correlated with the ECI calculated through the Method of Reflections, and more informative than the ECI in terms of predicting economic growth; the variables utilize ECI, PCI and export data to decompose a monolithic measure of complexity (ECI) into multiple dimensions, each capturing different information regarding the productive structure of the economy ($\rho$, $\nu$, number and complexity of capabilities). We thus conclude that the empirical success of ECI in predicting economic growth derives from its ability to condense multifaceted information present in the whole of a country's export distribution, particularly the quantity and quality of capabilities, into a single informative number. All results for this section are replicated for the years 2000, 2010 and 2015 in Appendix C. \\

To infer capabilities for each country in each year, we employ simulated annealing to search for an optimal set of capabilities which minimizes the KL divergence between the country's empirical export distribution against products complexity, smoothed and made continuous via a KDE (kernel density estimator) estimated using the improved Sheather-Jones method, and the model's predicted export distribution for the country given its set of capabilities. For technical details, see Appendix C.1. 

\subsection{From capabilities to economic development}

In this section, we will present details on the optimization of the model and show that the best values of $\rho$ and $\nu$ found for each economy varies robustly with both measures of economic complexity and interpret this result in terms of the productive structure of an economy. \\

The following table shows summary goodness-of-fit statistics for modelling empirical export distributions of 222 countries in 2005; in particular, "clarity" refers to the percentage reduction in KL divergence provided by our model compared to a uniform distribution representing maximum uncertainty, and can be interpreted as a $R^2$-like statistic representing the percentage of uncertainty in export distributions which can be explained by the model. 

\begin{table}[h]
\centering
\caption{Goodness-of-fit statistics for model on export distributions of 222 countries, 2005}
\begin{tabular}{lccc}
 \textbf{Statistic} & \textbf{Clarity}  & \textbf{KL Divergence} \\
\midrule
\textbf{Mean} & 0.174 & 1.574  \\ 
\textbf{St. Dev.} & 0.087 & 0.719 \\ 
\textbf{25\%} & 0.109 & 1.007 \\ 
\textbf{75\%} & 0.170 & 1.583
\end{tabular}
\end{table}

Using the explanatory variables in the previous section to proxy for economic development alongside ECI, its square, and the square of log GDP (centered about the mean), we fit an ordinal Logit model with the best values for $\rho$ and $\nu$ for different countries representing ordinal dependent variables ($\rho$ ordered from: $-\infty$ (Leontief), -9, -3, 0 (Cobb-Douglas), 1; $\nu$ ordered from: 0.5, 1, 2, 3, 4). A positive coefficient indicates a positive relationship with the probability of a country being in a higher category. \newpage

\begin{table}[h]
\centering
\caption{Logit regression for $\rho$ and $\nu$ against ECI, 2005}
\begin{tabular}{lcc}
\toprule
 & (1) & (2) \\
Variables & Logit regression for $\rho$ & Logit regression for $\nu$ \\
\midrule
Log GDP per capita & -0.109 &  -0.016 \\
 & (0.162) & (0.145) \\
Square log GDP per capita & -0.103 & -0.094\\
& (0.078) & (0.066) \\
Population & -0.004*** & 0.003* \\
& (0.001) & (0.001) \\
Investment-GDP ratio & 0.032 & 0.029* \\
& (0.019) & (0.016) \\
Export-GDP ratio &-0.010 & 0.009* \\
& (0.008) & (0.005) \\
ECI & 0.847** & -0.763*** \\
& (0.279) & (0.204)\\
ECI squared & 0.665** & 0.052 \\
& (0.203) & (0.123) \\
\midrule
Observations & 165 & 165 \\
(Psd.) R-squared & 0.142 & 0.075 \\
$\chi^2$-statistic, log-likelihood & 46.924*** & 36.736*** \\
AIC & 305.3 & 474.0 \\
\bottomrule
\end{tabular}
\begin{tablenotes}
\small
\item Notes: The dependent variable is different categories of $\rho$ and $\nu$.  Robust standard errors (HC1) are in parentheses. *** p$\textless$0.01, ** p$\textless$0.05, * p$\textless$0.1.
\end{tablenotes}
\end{table}

Identically-specified ordinal Logit regressions in 2000, 2010 and 2015 show a robust and significant positive relationship between ECI and $\rho$, of which a higher value indicates a higher elasticity of substitution $\sigma = \frac{1}{1-\rho}$ between capabilities in the model, as well as a robust and significant negative relationship between ECI and $\nu$, even after controlling for more conventional proxies of development such as log GDP per capita or its square. We also note that removing the ECI variables and leaving only log GDP per capita and its square results in a positive and significant coefficient for log GDP per capita for $\rho$, and a negative and significant coefficient for $\nu$, both of which are less statistically significant and yield a lower $R^2$ than with ECI alone. See Appendix C.3 for all preceding results. \\

Here, we note that these relationships - $\rho$ (and thus substitutability $\sigma = \frac{1}{1-\rho}$) increasing with GDP per capita (level of development), and $\nu$ decreasing - are prevalent throughout the bulk of the development economics literature (Chirinko 2008, Knoblach and Stockl 2019). In particular, $\rho$ increasing with GDP per capita indicates that more developed economies can combine capabilities more flexibly; and $\nu$ decreasing implies diminishing returns in combining capabilities, possibly because more complex products are closer to the technological frontier. The fact that ECI absorbs the significance of GDP per capita as a proxy for development in the regressions above is a very meaningful finding all on its own; it demonstrates that measures of complexity are more informative than aggregate measures like GDP per capita in determining the productive structure of an economy. As much of the literature has theoretically asserted via frameworks like the Solow model (Klump and de La Grandville 2000) and empirically confirmed (Miyagiwa and Papageorgiou 2002), a high rate of substitutability between different factors of production leads to a more efficient productive structure and a higher level of income; if ECI proxies for this measure of productive structure, then it would explain much of what makes ECI a robust predictor of economic growth in the first place.

\subsubsection{From capabilities to complexity}

Recall from Section 3 our zeroth-order and first-order measures for the complexity of a set of capabilities $A^*$:
\begin{equation}
    K_{A^*, 0} = |A^*|
\end{equation}
equalling the number of capabilities in $A^*$, and 
\begin{equation}
    K_{A^*, 1} = \frac{1}{K_{A^*, 0}} \sum_{a\in A^*} K_{a,1}
\end{equation}
equalling the average complexity of capabilities $a \in A^*$, which in turn are defined by 
\begin{equation}
    K_{a, 1} = \frac{1}{|P_{a\in p}|} \sum_{p \in P_{a \in p}}K_{p, 0}
\end{equation}
equalling the average complexity of products using capability $a$. Applied to an inferred set of capabilities $c^*$ for a real-world country, $K_{c^*, 0}$ and $K_{c^*, 1}$ become measures of economic complexity roughly analogous to diversity and ECI.  \\

Using these measures of complexity, this model will present two results which help further clarify the meaning behind ECI and what it measures. First, we will explore how these measures of economic development relate to one another: importantly, ECI correlates strongly with the quantity and quality of a country's capabilities. Second, as in Hidalgo and Hausmann's original paper introducing ECI (Hidalgo and Hausmann 2009), we conduct a series of Ordinary Least-Squares regressions of the derived complexity measures against average 20-year GDP per capita growth rates. We do not intend to portray our measure as an alternative to ECI; the chief aim is to establish that ECI can be interpreted in terms of capabilities (as well as the productive structure of an economy), and that the latent capabilities of countries inferred above provide the same information as the ECI. Results for different starting years (2000, 2010, 2015) will be included in Appendix C.4. \\

To begin with, we observe that these measures are significantly - but not perfectly - correlated with one another. Of particular interest is the high correlation ($>0.5$) between all three measures of complexity (ECI, number of capabilities, average capability complexity) and moderate correlation between these measures of complexity and diversity. Even more interestingly, we note that average capability complexity is not particularly correlated with log GDP per capita compared with ECI (0.27 vs. 0.65). This suggests not only that these measures of complexity capture different but related realms of information regarding economic growth, but also that their relationship is not purely due to both being related to a traditional variable of economic development (GDP per capita) and instead points towards some intrinsic link between ECI and the quantity and quality of a country's capabilities. 

\begin{center}
    \begin{figure}[h]
        \includegraphics[width=12cm]{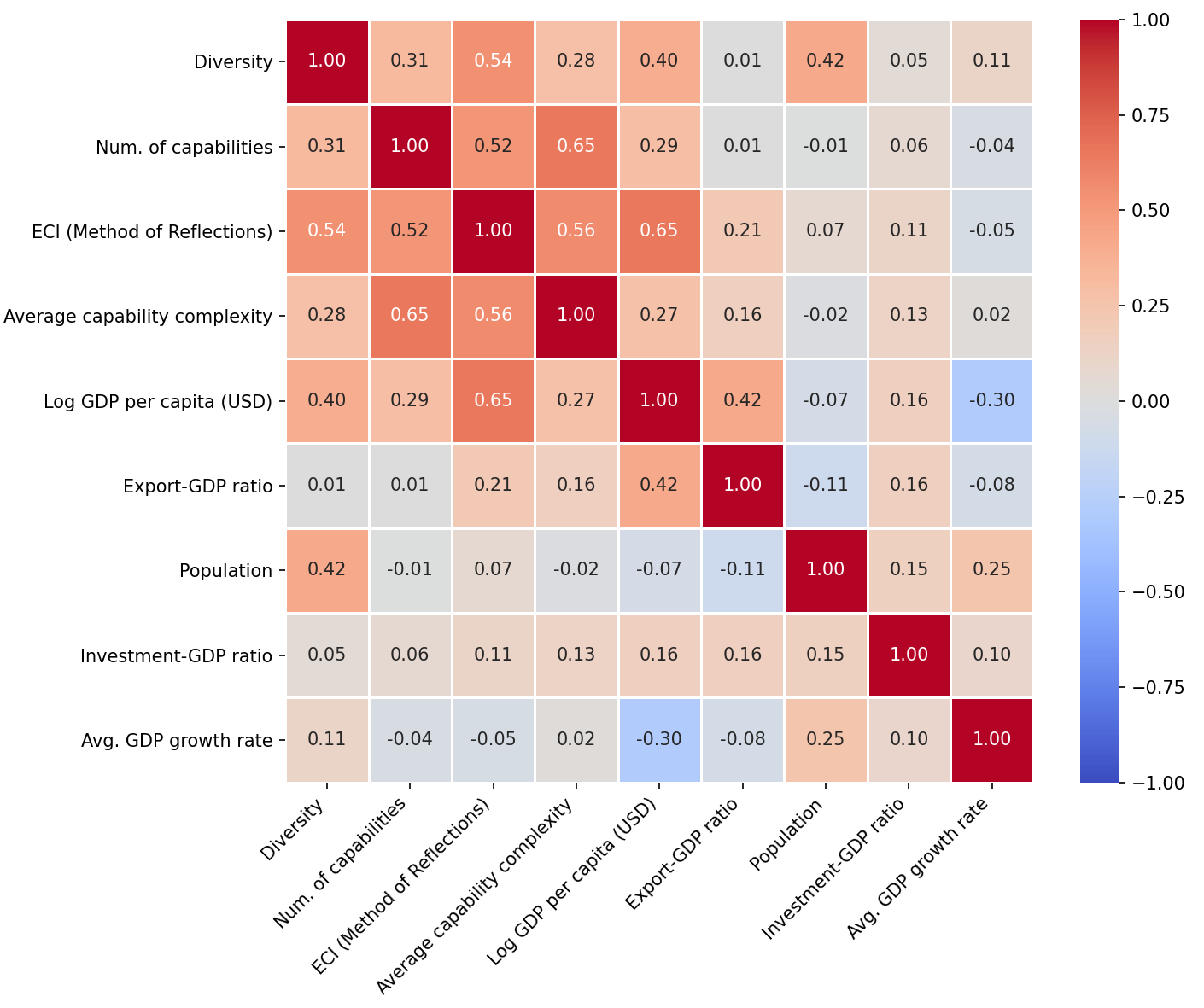}
        \caption{Heatmap of correlations between measures of economic development}
    \end{figure}
\end{center}

To conduct regressions against growth, we eliminate all observations with missing data from our dataset, leading to 160 remaining countries; the regression specification is Ordinary Least Squares (OLS) with HC1 (heteroskedasticity-robust) covariance and no multicollinearity detected by the Variance Inflation Factor (VIF) test, with VIF of all variables below 10. We conduct the following regressions in order:

\begin{enumerate}
    \item Original ECI and log GDP per capita, as in Hidalgo and Hausmann's original paper.
    \item Original ECI, log GDP per capita, investment-GDP ratio, export-GDP ratio.
    \item Average capability complexity and log GDP per capita.
    \item Average capability complexity, log GDP per capita, investment-GDP ratio, export-GDP ratio.
\end{enumerate}

The main body of this paper will present only the second and fourth regressions, as a means of comparing ECI and average capability complexity under the most stringent set of explanatory variables (note population was not included due to multicollinearity); for results from the other two regressions, see Appendix C.4.

\begin{table}[h]
\centering
\caption{Regression for ECI vs. average capability complexity against 20-year average growth, 2005}
\begin{tabular}{l c c}
\toprule
 & \multicolumn{2}{c}{Growth, 20-year average} \\
\cmidrule(lr){2-3}
 & (1) ECI & (2) Capability complexity \\
\midrule
ECI & 0.394* & \\
 & (0.202) & \\
Average capability complexity & & 0.285** \\
 & & (0.140) \\
Log GDP per capita & -0.617*** & -0.463*** \\
 & (0.136) & (0.084) \\
Investment-GDP ratio & 0.021 & 0.011 \\
 & (0.016) & (0.015) \\
Export-GDP ratio & 0.003 & 0.004 \\
 & (0.004) & (0.005) \\
Population & & 0.003** \\
 & & (0.001) \\
Constant & 6.439*** & 5.230*** \\
 & (1.141) & (0.800) \\
\midrule
Observations & 165 & 165 \\
R-squared & 0.107 & 0.151 \\
AIC & 669.5 & 662.0 \\
\bottomrule
\end{tabular}
\medskip
{\small \textit{Notes:} Robust standard errors (HC1) in parentheses. *** p$\textless$0.01, ** p$\textless$0.05, * p$\textless$0.1. The capability model includes population as an additional control.}
\end{table}

In summary, we observe a similar adjusted and unadjusted $R^2$ when including average capability complexity (0.151) compared to ECI (0.107), as well as a lower AIC (662 compared to 669.5); while both variables are significant when controlling for log GDP per capita, average capability complexity remains significant at a 5\% significance level even with the inclusion of all explanatory variables. 
\section{Discussion and conclusion}

This paper makes contributions to three foundational building blocks of the economic complexity literature: the Product Space, the ECI, and that which ties both pillars of economic complexity theory together - the study of capabilities. The core of our contribution is our model of the Capability Space - a latent network of related capabilities that combine to make products and give productive structure to countries; this capability-centric perspective on economic complexity reframes relationships between products in the Product Space as interconnections between capabilities, PCI and ECI as measures of capability quantity and quality, and complexity as a whole as a dimensionality-reducing encapsulation of a country's productive structure containing the seeds for future growth. \\

In terms of the Product Space in particular, and the study of complex networks in general, this paper represents the first comprehensive study of the complex-network properties of the Product Space beyond its basic macroscopic structure (whose core-periphery and community properties have previously been noted). We find that the Product Space is characterized by several highly distinctive complex-network properties, including an extremely left-skewed degree distribution and right-skewed weight distribution where the majority of products are connected, but few are connected strongly; furthermore, the Product Space is mathematically and empirically underpinned by the Product Complexity Index, in which products of similar complexities are more strongly connected and more complex products occupy more central positions in the network. \\

In Section 5, our capability-centric model sheds more light on the relationships between products evident in the Product Space by introducing a plausible mechanism for explaining why products are related: by interpreting products as combinations of latent capabilities which accrue through a mechanism of modified preferential attachment governed by the underlying Capability Space (assumed to be in block-matrix form), and interpreting PCI as the number of capabilities required to produce a product, the model reproduces the key global topological properties of the Product Space, including its relationship to PCI, and its macroscopic community and core-periphery structure. It is worth noting that this preferential attachment mechanism driven by a latent network is broadly generalizable. For example, an analogy could easily be drawn to networks such as the Research Space (Guevara et al. 2016), which connects pairs of research areas to one another based on shared publications, or social networks connecting individuals on social media platforms; what could be a model of capabilities combining to form products could just as easily be a model of research foci combining to form papers, or hobbies and interests combining to form social media profiles. These networks present vastly different topologies and properties compared to the Product Space; the Research Space, for example, is characterized by a "ring-like" topology comprised of multiple peripheral clusters surrounding a core cluster. But if "birds of a feather" truly flock together in social networks - if nodes comprised of similar capabilities, similar research foci, and similar interests connect to one another more strongly, as research on the principle of homophily would suggest (McPherson et al. 2001) - then this model draws on that principle to the fullest. \\ 

The key empirical finding of the paper, however, lies in its re-interpretation of the ECI through the lens of the Capability Space in Section 6. With Section 5 solidifying the Capability Space as a plausible model for the topology of the Product Space, Section 6 uses a CES-inspired production function that arises from a natural generalization of the original Leontief production function used by Hidalgo and Hausmann to model the export distribution of countries given a set of latent capabilities; it thus proposes that the same capability-centric view of products is applicable to countries, not as a predictive model but as an explanatory mechanism for what the ECI measures. Most saliently, we obtain estimates for crucial parameters in the CES production function for countries by finding the optimal set of capabilities that best explain countries' empirical export distributions. We find that $\rho$, the parameter controlling for the substitutability of capabilities, and $\nu$, the returns-to-scale parameter, obtained by the model for countries at different levels of development are robustly linked to measures of complexity; importantly $\rho$ increases with complexity after controlling for GDP per capita, population, etc. This provides a powerful mechanism for understanding why economic complexity robustly predicts future economic growth: while diminishing returns-to-scale will near-inevitably occur at the middle-to-high income transition, a higher economic complexity means a more flexible productive structure in which different capabilities are fluid and easily substitutable. Furthermore, we find that the quantity and quality of capabilities that best explain countries' empirical export distributions - encapsulated by the number of capabilities countries are endowed with (quantity) and the average complexity of the products they produce (quality) - are highly correlated with the ECI of countries, and capture similar information when used in strictly-specified OLS regressions against 20-year growth. \\

In effect, the ECI is an empirically successful proxy for the scope and extent of these capabilities - not just an empirically successful forecaster of economic growth. This conclusion is remarkable not only in its ability to reconcile the longstanding tension between economic complexity and the model of capabilities it wields, but also in the fact that that it was reached theoretically in near-concurrent work by Hidalgo and Stojkoski. In their paper, Hidalgo and Stojkoski analytically solve the original Hidalgo-Hausmann model of capabilities under the same realization that an output function, rather than a binary specialization function, was needed to define ECI under the scope of the model. Through an analytical solution for a simpler single-capability variant of the model and a numerical simulation for a multi-capability model of combinatorial capabilities, they prove a mathematical interpretation of the ECI eigenvector as a perfect estimator of "the average capability endowment of an economy", dividing the world into economies which are most likely to possess a specific capability and economies which are least likely to do so. Though our work tackles an identical research problem, our contributions are fundamentally complementary - even mutually reinforcing - in nature. \\

Most importantly, the two papers diverge in our approach to extending the original Hidalgo-Hausmann model of combinatorial capabilities. Though both papers culminate in a production function that link countries to output levels of specific products through capabilities, their work incorporates this by differentiating capabilities through the lens of their ownership by countries or products: a (possibly varying) parameter is used to control for the probability that a country possesses, or product requires, a certain capability, translating directly into a production level that varies with such. Meanwhile, our work differentiates capabilities through treating capabilities themselves as interrelated; it is capability-centric in the sense that countries or products are not assumed to be described by any attributes aside from the set of capabilities that determine them, and the emphasis is on the capabilities themsleves and the relationships between them. In addition, one of the most significant contributions of Hidalgo and Stojkoski was that their analytical derivation of ECI as an estimator of capability ownership could generalize to any non-multiplicatively separable production function. With a single exception for $\rho = 0$ (the Cobb-Douglas production function), the CES production function we generalize from the original model's Leontief production function satisfies this condition of non-separability (see Appendix A.2 for proof), and as such allows the empirical results of this paper in correlating ECI with an explicit measure of capability endowment to benefit from the rich theoretical underpinnings of Hidalgo and Stojkoski's contributions. \\

Of course, it is worth noting that the empirical results presented in this paper are also an empirical validation of the core concept of both papers' methods; while the numerical simulation of the multi-capability model presented by Hidalgo and Stojkoski proved to be robust against substantial noise, we validate their central premise and show that a model of related capabilities forms the crux of the underlying mechanism behind both the topology of the Product Space and the shape of countries' export distributions. This paper does not claim to be a thorough computational investigation of the Capability Space model, merely a proof-of-concept of a capability-centric model of economic complexity: in particular, genuine limitations on the results presented here include a failure to match the empirical centrality distribution of the Product Space (Section 5) and significant room for improvement in optimizing country capability sets through simulated annealing, as computational constraints meant we could not guarantee a global minimum or a uniquely optimal solution. These aspects could prove to be fertile ground for future research in the area. \\

Underscored by these new developments in bridging complexity to capabilities, future work in the field of economic complexity is rife with possibility. Aside from applying the model presented in this paper to the context of similar networks of relatedness with entirely distinct macroscopic structures, such as the Research Space, the methods in this paper could be expanded along dynamic lines; as the model provides a method to infer capabilities from export baskets, a natural follow-up question to ask is how these capabilities could evolve over time - for instance, would a dynamic simulation of economic development through treating the Product Space as a transition matrix between products be able to tell us anything about capability acquisition or replicate longstanding results foundational to studies in economic development? Numerical simulations of the model with massively increased granularity are also possible - increasing the number of capabilities in the model or making $\rho$ and $\nu$ continuous, for example, is very likely to yield measures of capability complexity that are more informative than ours in terms of predicting economic growth. Further explorations of the model along these lines could yield fruitful results.  \\

For half a century, development economics has evolved by emancipating the notion of economic capabilities from its black box of exogeneity; and as advances in network science continue to mature, and economic data reaches further to encompass cities, towns, and ever-finer sub-divisions of products, so, too, will the latest and most daring journey into this black box -  the field of economic complexity - emancipate capabilities from the realm of the unobservable towards the surface-world of economics at last.

\subsection{Code and data availability}

All code and data files used in this paper are available online at \url{https://github.com/Aurore32/Across-Time-and-Product-Space}.

\clearpage
\nocite{*}
\bibliography{paper}
\clearpage

\begin{appendices}
\setcounter{figure}{6} 
\renewcommand{\thefigure}{\arabic{figure}}
\setcounter{table}{8} 
\renewcommand{\thetable}{\arabic{table}}

\section{Additional results for Section 4: A model of related capabilities}

\subsection{Additional results for Section 4.1: Capability blocks}

\begin{center}
    \begin{figure}[h]
        \includegraphics[width=13cm]{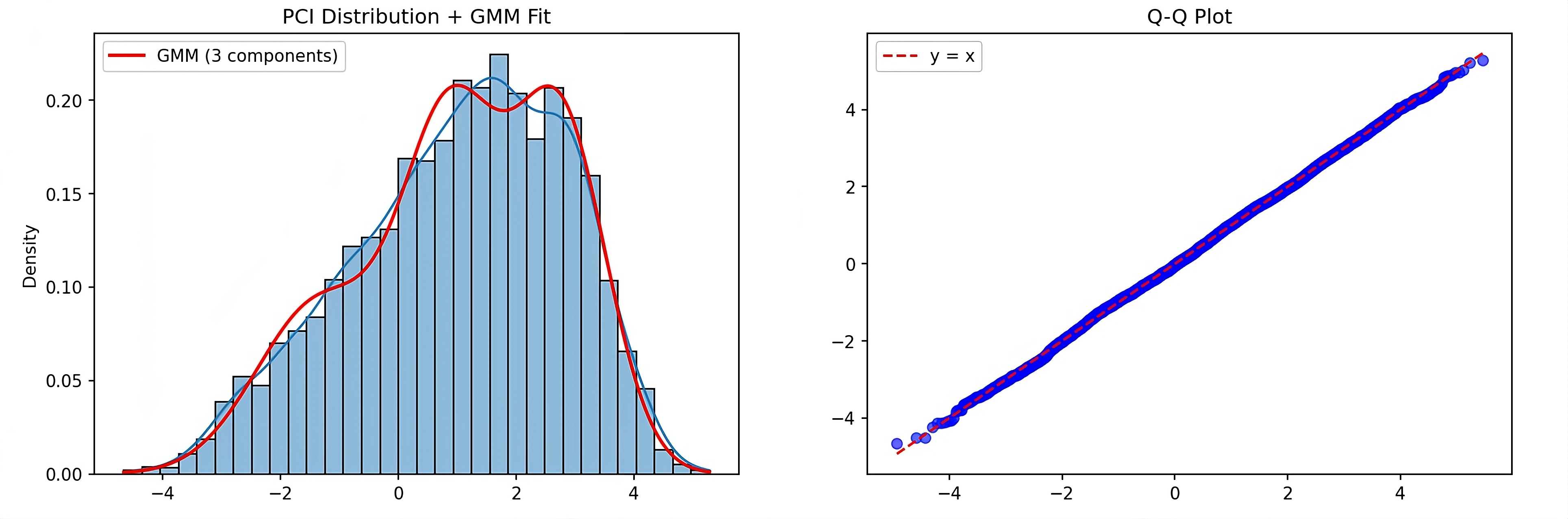}
    \caption{A 3-component GMM fitted onto the empirical PCI distribution of 5008 products in 2005. Note the visible goodness-of-fit of the GMM to the data, as well as the closeness of the Q-Q (quantile-quantile) plot to the 45-degree line, which indicates equivalence between the quantiles of the two distributions}
    \end{figure}
\end{center}

The AIC (Akaike information criterion) was used to select the GMM which represented the best model for the PCI distribution. A test using BIC instead of AIC shows the same result: that a 2-component GMM achieves the best goodness of fit while maximizing parsimony. Higher-component GMMs achieve marginally better goodness-of-fit, but are excluded from the following table because the two-component GMM is already statistically indistinguishable from the PCI distribution.

\begin{table}[h]
\caption{Goodness-of-fit of GMMs on empirical PCI data}
\begin{tabular}{lcccc}

                                                                & \textbf{Number of components with min. AIC} & \textbf{p-value for KS Test}\\
\midrule
\textbf{2000} & 2 & 0.360* \\
\textbf{2005} & 2 & 0.669* \\
\textbf{2010} & 2 & 0.312* \\
\textbf{2015} & 2 & 0.375*
\end{tabular}
\begin{tablenotes}
\small
\item (*a $p$-value above 0.05 indicates non-rejection of the null hypothesis, which suggests the two samples being compared come from the same underlying distribution. All tests were conducted by comparing PCI of 5008 products using the HS 6-digit classification vs. the calcuated CDF of the GMM.)

\end{tablenotes}
\end{table}

\subsection{Additional results for Section 4.4: The production function}
Recall the generalized form of our production function derived in Section 4.4:

\begin{equation}
Q(c,p) = \alpha(\sum_{i=1}^{K_{p,0}} \frac{1}{K_{p,0}} \phi(c,\{p_i\})^\rho)^\frac{\nu}{\rho}
\end{equation}

This generalized production function embeds within itself other production functions that have previously been used to model capabilities; in particular, the Leontief-like production function of the original Hidalgo-Hausmann model is recoverable from this generalized form when $\rho$ approaches $-\infty$, $\nu$ is one and $\sigma$ approaches zero, meaning that capabilities are not at all substitutable. Denote $\min\{\phi(c,\{p_i\}),\ i=1, 2, ..., K_{p,0}\}$, representing the capability $p_i \in p$ least related to $c$, as $\phi^* = \phi(c,\{p_i^*\})$, where $1 \leq i^* \leq K_{p,0}$; then we have

\begin{equation}
    \begin{aligned}
        \lim_{\rho \to -\infty} \alpha(\sum_{i=1}^{K_{p,0}} \frac{1}{K_{p,0}} \phi(c,\{p_i\})^\rho)^\frac{1}{\rho} &= \lim_{\rho \to -\infty} \alpha\phi^*(\sum_{i=1}^{K_{p,0}} \frac{1}{K_{p,0}}  (\frac{\phi(c,\{p_i\})}{\phi^*})^\rho)^\frac{1}{\rho} \\ 
    \end{aligned}
\end{equation}
As $\phi^* \leq \phi(c,\{p_i\})$ for any $i = 1, 2, ..., K_{p,0}$ by definition, the limiting value as $\rho \to -\infty$ of $(\frac{\phi(c,\{p_i\})}{\phi^*})^\rho$ is either zero (if the numerator is strictly greater than the denominator) or one (if the two are equal). Thus the value $\sum_{i=1}^{K_{p,0}} \frac{1}{K_{p,0}}  (\frac{\phi(c,\{p_i\})}{\phi^*})^\rho$ converges to a finite positive number; and raising that number to the power $\frac{1}{\rho} \to 0$ gives exactly one. Therefore, the limiting value of this production function as $\rho \to -\infty$ is simply

\begin{equation}
    \lim_{\rho \to -\infty} Q(c,p) = \alpha \phi^*
\end{equation}
where $\phi^*$ represents the capability least related to the current capabilities of $c$. When the proximity function $\phi(c, \{p_i\})$ is chosen to be the maximum value instead of the average value of capability proximity $\Phi^C_{cp_i}$ from the capabilities possessed by $c$ to the capability $p_i$, and we assume a diagonal Capability Space where all capabilities are only related to themselves with relatedness $1$ and unrelated to all others, this limiting case is identical to the Hidalgo-Hausmann production function as it mandates that $\phi^*$ equals $1$ for production to occur. \\

Additionally, we note that this production function is non-multiplicatively separable and thus satisfies the condition given by Hidalgo and Stojkoski  for interpreting an ECI derived from such a production function as a perfect estimator of an economy's capability endowment. We proceed via proof by contradiction. Suppose that $Q(c,p)$ was multiplicatively separable in each of the inputs $\phi(c, \{p_i\})$, denoted $\phi_i$ for simplicity; then we would have 
\begin{equation}
    Q(c,p) = f_1(\phi_1) f_2(\phi_2) ... f_{K_{p,0}}(\phi_{K_{p,0}}).
\end{equation}
If a function is multiplicatively separable as above, then its logarithm must be additively separable in the logarithms of each of $f_1, f_2, ..., f_{K_{p,0}}$, i.e.
\begin{equation}
    \log Q(c,p) = \log f_1(\phi_1) + \log f_2(\phi_2) + ... + \log f_{K_{p,0}}(\phi_{K_{p,0}})
\end{equation}
with also
\begin{equation}
    \log Q(c,p) = \log \alpha + \frac{\nu}{\rho} \log \sum_{i=1}^{K_{p,0}} \frac{1}{K_{p,0}} \phi_i^\rho
\end{equation}
The simplest way to check for separability is through the partial derivative; if the equation above holds, then taking 
\begin{equation}
    \frac{\partial}{\partial \phi_i} Q(c,p)
\end{equation}
for any $i=1, 2, ..., K_{p,0}$ would result simply in $\frac{\partial}{\partial \phi_i} = \frac{f_i'(\phi_i)}{f_i (\phi_i)}$ as the other terms do not contain $\phi_i$ (by definition of additive separability), where $f_i'$ denotes differentiation by $\phi_i$. As such, we have 
\begin{equation}
\begin{aligned}
    \frac{\partial}{\partial \phi_i} \log Q(c,p) &= \frac{f_i'(\phi_i)}{f_i (\phi_i)} \\
    \frac{\partial}{\partial \phi_i} (\log \alpha + \frac{\nu}{\rho} \log \sum_{i=1}^{K_{p,0}} \frac{1}{K_{p,0}} \phi_i^\rho) &= \frac{f_i'(\phi_i)}{f_i (\phi_i)} \\
    \frac{\nu}{\rho}\frac{1}{K_{p,0}}\frac{\rho\phi_i^{\rho-1}}{\sum_{i=1}^{K_{p,0}} \frac{1}{K_{p,0}} \phi_i^\rho} &= \frac{f_i'(\phi_i)}{f_i (\phi_i)} \\
\end{aligned}
\end{equation}
where no equivalence can be drawn because the left-hand side depends on all of $\phi_1, \phi_2, ..., \phi_{K_{p,0}}$, while the right-hand side is only a function of $\phi_i$. $\blacksquare$

\section{Additional results for Section 5: Modeling the Product Space}

\subsection{Additional results for Section 5.1: Topology of the Product Space}

\begin{table}[h]
\centering
\caption{Summary statistics of the Product Space, 2000-2015}
\begin{tabular}{lcccc}

                                                                & \textbf{Number of nodes} & \textbf{Density} & \textbf{Clustering coefficient}  & \textbf{Modularity}\\
\midrule
\textbf{2000}                                                    &  5017 & 0.913     &   0.931     &  0.114  \\ 
\textbf{2010}                                                    &  4938 & 0.896    &   0.927    &  0.111  \\ 
\textbf{2015}                                                    &  4865 & 0.881     &   0.918     &  0.102  \\ 
\end{tabular}
\end{table}

There is an observed downward trend in the density and clustering coefficient of the network that seems to be suggestive of increasing sparsity and a less connected Product Space, but the fluctuations in the summary statistics are not at all dramatic. \\

In addition, we present the following results for edge weight, centrality, and degree distributions across the years:

\begin{table}[h]
\centering
\caption{Distribution statistics for the Product Space, 2000–2015}
\begin{tabular}{l c c c}
\toprule
\multirow{2}{*}{Statistic} & \multicolumn{3}{c}{Year} \\
\cmidrule(lr){2-4}
 & 2000 & 2010 & 2015 \\
\midrule
\multicolumn{4}{l}{\textbf{Centrality}} \\
Mean & 0.571 & 0.583 & 0.574 \\
Median & 0.585 & 0.604 & 0.596 \\
Mode & 0.004* & 0.047* & 0.030* \\
IQR & 0.216 & 0.245 & 0.249 \\
Skew & -0.457 & -0.589 & -0.574 \\
Kurtosis & -0.063 & 0.064 & 0.005 \\
\midrule
\multicolumn{4}{l}{\textbf{Weight}} \\
Mean & 0.156 & 0.153 & 0.152 \\
Median & 0.143 & 0.136 & 0.135 \\
Mode & 0.167 & 0.167 & 0.167 \\
IQR & 0.127 & 0.124 & 0.119 \\
Skew & 0.941 & 0.964 & 1.01 \\
Kurtosis & 0.984 & 1.10 & 1.28 \\
\midrule
\multicolumn{4}{l}{\textbf{Degree}} \\
Mean & 4579 & 4426 & 4285 \\
Median & 4727 & 4606 & 4488 \\
Mode & 4885 & 1180 & 670 \\
IQR & 4850 & 4739 & 4629 \\
Skew & -3.04 & -3.39 & -3.12 \\
Kurtosis & 14.1 & 14.1 & 12.1 \\
\bottomrule
\end{tabular}
\medskip
{\small \textit{Notes:} Mode is not informative for centrality because the distribution is continuous; the reported values are isolated occurrences. Degree scales differ slightly across years (maximum degree ~4900±100); raw values are shown.}
\end{table}

\subsection{Additional results for Section 5.2: Product complexity in the Product Space}
\newpage
\begin{table}[h]
\centering
\caption{Product complexity in the Product Space, 2000–2015}
\begin{tabular}{l c c c}
\toprule
 & \multicolumn{3}{c}{Year} \\
\cmidrule(lr){2-4}
 & 2000 & 2010 & 2015 \\
\midrule
PCI–centrality correlation & 0.328* & 0.305 & 0.326 \\
$\Delta$PCI–proximity correlation & -0.377 & -0.286 & -0.271 \\
PCI modularity & 0.079 & 0.071 & 0.066 \\
\bottomrule
\end{tabular}
\medskip
{\small \textit{Notes:} *All correlation coefficients are statistically significant at the 1\% level.}
\end{table}

\subsection{Additional results for Section 5.3: Results for the model}

The following section will present two supplementary results from a simulation of the model on the Product Space. First, the following tables represent the tuned parameters during model optimization as well as the values converged upon through CMA-ES for simulation of the Product Space in the year 2005 presented in the main body of the paper, respectively:

\begin{table}[h]
\centering
\caption{Parameter summary for Product Space simulation}
\begin{tabular}{lcccc}

 \textbf{Parameter} & \textbf{Parameter name}  & \textbf{Tuned?} & \textbf{Range} & \textbf{Initial value} \\
\midrule
$N_p = \kappa$ & Number of products & No & 1000 & 1000 \\
n/a & $\Phi^C$ distribution type & No & Constant, Beta & Constant \\
  $C$ & Max capabilities for a product  & No & 100 & 100 \\
 $C_{c}$ & Size of core block  & No & 25 & 25 \\ 
 $C_{p}$ & Size of periphery block  & No & 25  & 25 \\ 
  $n$ & Number of GMM components & No & 8 & 8 \\

 $\phi^{p}_{b}$ & Proximity between periphery blocks  & Yes & 0.01-0.1  & 0.05 \\ 
 $\phi^{p}_{w}$ & Proximity within periphery blocks  & Yes & 0.8-0.99  & 0.9 \\ 
  $\phi^{p}_{c}$ & Proximity from periphery to core blocks  & Yes & 0.01-0.1  & 0.05 \\ 
 $\phi^{c}_{b}$ & Proximity between core blocks  & Yes & 0.01-0.1  & 0.05 \\ 
 $\phi^{c}_{w}$ & Proximity within core blocks  & Yes & 0.8-0.99  & 0.9 \\
  $\phi^{c}_{p}$ & Proximity from core to periphery blocks  & Yes & 0.01-0.1  & 0.05 \\ 
 
\end{tabular}
\end{table}

\newpage
\begin{table}[h]
\centering
\caption{Parameter results for Product Space simulation}
\begin{tabular}{lccc}
 \textbf{Parameter} & \textbf{Parameter name}  & \textbf{Parameter value} \\
\midrule
$\phi^{p}_{b}$ & Proximity between periphery blocks  & 0.096 \\ 
$\phi^{p}_{w}$ & Proximity within periphery blocks   & 0.990 \\ 
$\phi^{p}_{c}$ & Proximity from periphery to core blocks   & 0.010 \\ 
$\phi^{c}_{b}$ & Proximity between core blocks    & 0.010 \\ 
$\phi^{c}_{w}$ & Proximity within core blocks  & 0.989 \\ 
$\phi^{c}_{p}$ & Proximity from core to periphery blocks  & 0.018 \\ 
\end{tabular}
\end{table}

Furthermore, we conduct sensitivity analyses of the Product Space simulation to the following parameters, adjusted one at a time. Aside from the first listed test, all subsequent tests will be conducted on the model for 2005 data only; note that test 3 (changing the number of GMM components) alters the core structure of the model and thus requires parameter re-tuning. All CMA-ES runs are conducted with population size $20$ and a single optimization run of $50$ generations, choosing the parameters which minimize the KS distance to the empirical weight distribution. Note that we do not alter the search space of the parameters to ensure comparability of results.
\begin{enumerate}
    \item The year used for empirical data (2000, 2005, 2010, 2015), using a constant distribution for $\Phi^C$ for each.
    \item The distribution for $\Phi^C$ used. In the main body of the paper, we used constant weights within and between blocks; an alternative is using the Beta distribution
\begin{equation}
    f(\Phi^C_{ij}) = \frac{1}{B(\alpha, \beta)}(\Phi^C_{ij})^{\beta - 1}(1+\Phi_{ij}^C)^{-\alpha - \beta}
\end{equation}
where $B(\alpha, \beta)$ is a normalizing parameter explicitly defined via the Gamma function, and
\begin{equation}
    \begin{cases}
        \alpha = \kappa \cdot \mu_{ij} \\
        \beta = \kappa\cdot (1-\mu_{ij})
    \end{cases}
\end{equation}
where the mean of this distribution is $\mu_{ij}$, the origianl constant value of the weight $\Phi^C_{ij}$ in the Capability Space, and $\kappa$ is a scale parameter. \\

    In effect, this distribution introduces random noise into the Capability Space such that all parameters are heterogeneous, even within blocks. This tests whether heterogeneity between capabilities is at all required to capture a more intricate topology of the Product Space.
    \item The value of $n$ (number of GMM components, equalling the number of blocks) used. As this fundamentally changes the structure of the Capability Space, the parameters to the model are re-tuned via CMA-ES using a warm-start with initial conditions equal to the previous optimum presented in the main body of the paper. Values: $2$, $4$, $6$, $8$, $10$ (original is $8$). 
    \item The value of $\kappa$ for the Beta distribution, representing heterogeneity of capabilities within a block. Parameters will not be re-tuned; only $\kappa$ will be changed. Lower $\kappa$ leads to greater heterogeneity. Values: $200, 400, 600, 800, 1000$ (original is $1000$).
\end{enumerate}
\subsubsection{Results for multiple years}
The following present Test 1: results from multiple years, with the original parameters left as-is and the block matrix elements re-tuned via CMA-ES. The results between all four years show a striking similarity in the parameters found to be optimal (higher between-periphery proximity than within-core proximity, suggestive of a less specialized periphery), as well as their ability to reproduce the core topological properties of the Product Space. \\

\begin{table}[h]
\centering
\caption{Product Space simulation parameters for 2000, 2010, and 2015}
\begin{tabular}{l c c c}
\toprule
\multirow{2}{*}{Parameter} & \multicolumn{3}{c}{Year} \\
\cmidrule(lr){2-4}
 & 2000 & 2010 & 2015 \\
\midrule
$\phi^{p}_{b}$ (periphery–periphery) & 0.098 & 0.099 & 0.100 \\
$\phi^{p}_{w}$ (within periphery) & 0.988 & 0.989 & 0.981 \\
$\phi^{p}_{c}$ (periphery → core) & 0.010 & 0.011 & 0.010 \\
$\phi^{c}_{b}$ (core–core) & 0.010 & 0.010 & 0.010 \\
$\phi^{c}_{w}$ (within core) & 0.988 & 0.990 & 0.989 \\
$\phi^{c}_{p}$ (core → periphery) & 0.017 & 0.010 & 0.011 \\
\bottomrule
\end{tabular}
\medskip
{\small \textit{Notes:} Parameters were tuned separately for each year using CMA-ES.}
\end{table}

\begin{table}[h]
\centering
\caption{Product Space simulation results for 2000, 2010, and 2015}
\begin{tabular}{l c c c c c c}
\toprule
\multirow{2}{*}{Property} & \multicolumn{2}{c}{2000} & \multicolumn{2}{c}{2010} & \multicolumn{2}{c}{2015} \\
\cmidrule(lr){2-3} \cmidrule(lr){4-5} \cmidrule(lr){6-7}
 & Empirical & Simulated & Empirical & Simulated & Empirical & Simulated \\
\midrule
Modularity (Leiden) & 0.114 & 0.134 & 0.111 & 0.134 & 0.102 & 0.125 \\
PCI modularity & 0.079 & 0.067 & 0.071 & 0.061 & 0.066 & 0.068 \\
$\Delta$PCI–proximity correlation & -0.377 & -0.401 & -0.286 & -0.421 & -0.271 & -0.412 \\
PCI–centrality correlation & 0.344 & 0.284 & 0.305 & 0.254 & 0.326 & 0.249 \\
Degree distribution (KS D) & --- & 0.059 & --- & 0.113 & --- & 0.174 \\
Weight distribution (KS D) & --- & 0.044* & --- & 0.047* & --- & 0.058* \\
Centrality distribution (KS D) & --- & 0.282 & --- & 0.454 & --- & 0.356 \\
\bottomrule
\end{tabular}
\medskip
{\small \textit{Notes:} KS D‑statistics are for the goodness‑of‑fit test between simulated and empirical distributions. * indicates that the null hypothesis of identical distributions cannot be rejected at the 5\% significance level (D $<$ 0.061).}
\end{table}

\subsubsection{Results for the Beta distribution and GMM components}
The following present Test 2: the use of a Beta distribution, and Test 3: varying the number of GMM components, using data from 2005 and $\kappa = 1000$ (equal to the number of products). \\ 

The Beta distribution essentially serves as an introduction of random noise into the Capability Space; given these conditions, we note that the Beta distribution provides a better fit for the centrality distribution than the constant block matrix and is just as effective at reproducing the other properties below. Curiously, we note that this goodness-of-fit is not particularly sensitive to $\kappa$; this may suggest that the topology of the Product Space does not necessarily depend on the individual differences between capabilities as much as it depends on the broader community structure of capabilities, as suggested by the block matrix model. The table below provides an overview of the tuned parameters for both the Beta distribution and for different GMM components; parameters are kept the same for all tests involving the Beta distribution (though $\kappa$ varies), and re-tuned for varying numbers of GMM components. For varying GMM components, no other parameters are adjusted, except $C_c$ and $C_p$, the number of capabilities in each block; this is strictly done via $n \times C_c = n \times C_p = 200$ as in the original model with $n = 8, C_c = 25$. This will mean that the ratio between $C_c$ and the maximum number of capabilities a product can require will also be adjusted. \\

As shown below, the Beta distribution does not alter the efficacy of the model in replicating the topology of the Product Space. However, all specifications for the model below $6$ components fail at replicating most of the Product Space's emergent properties accurately, including the degree and centrality distributions, and overestimate the community structure of the network because there are simply too few capability blocks to form a wide variety of distinct product clusters. This may suggest a more granular picture of capabilities (e.g. occupations), with a large number of clusters that represent capabilities related strongly to one another and weakly to capabilities in other clusters, in the real world. \\

The model is slightly sensitive to the value of $n$ because it fundamentally alters the structure of the Capability Space. A lower value of $n$ (2, 4) leads to an overly modular Product Space, with a very clear dichotomy between low- and high-complexity products; a high value of $n$ leads to more ability to replicate the underlying distributions (centrality, weight, degree) but an unrealistically granular Product Space with low modularity and weak core-periphery structure (low centrality correlation with PCI). This should be interpreted more as commentary on the underlying structure of the Product Space than as a deficiency of the model; different networks have different levels of modularity, which necessitate different levels of granularity for the Capability Space. \newpage

\begin{table}[h]
\centering
\caption{Tuned Capability Space parameters for sensitivity analyses}
\begin{tabular}{l c c c c c c}
\toprule
\multirow{2}{*}{Scenario} & \multicolumn{6}{c}{Parameter} \\
\cmidrule(lr){2-7}
 & $\phi^{p}_{b}$ & $\phi^{p}_{w}$ & $\phi^{p}_{c}$ & $\phi^{c}_{b}$ & $\phi^{c}_{w}$ & $\phi^{c}_{p}$ \\
\midrule
Beta distribution ($\kappa=1000$) & 0.095 & 0.982 & 0.010 & 0.010 & 0.973 & 0.023 \\
\midrule
\multicolumn{7}{l}{Number of GMM components} \\
2 components & 0.068 & 0.837 & 0.029 & 0.010 & 0.816 & 0.063 \\
4 components & 0.010 & 0.802 & 0.047 & 0.010 & 0.802 & 0.014 \\
6 components & 0.063 & 0.800 & 0.013 & 0.012 & 0.800 & 0.012 \\
8 components & --- & --- & --- & --- & --- & --- \\
10 components & 0.100 & 0.989 & 0.010 & 0.010 & 0.970 & 0.017 \\
\bottomrule
\end{tabular}
\medskip
{\small \textit{Notes:} The 8‑component case is the baseline model used in the main text (parameters reported in Table 16). All parameters were tuned via CMA-ES. $\phi^{p}_{b}$: periphery–periphery; $\phi^{p}_{w}$: within periphery; $\phi^{p}_{c}$: periphery→core; $\phi^{c}_{b}$: core–core; $\phi^{c}_{w}$: within core; $\phi^{c}_{p}$: core→periphery.}
\end{table}

\begin{center}
    \begin{figure}[h!]
        \includegraphics[width=15cm]{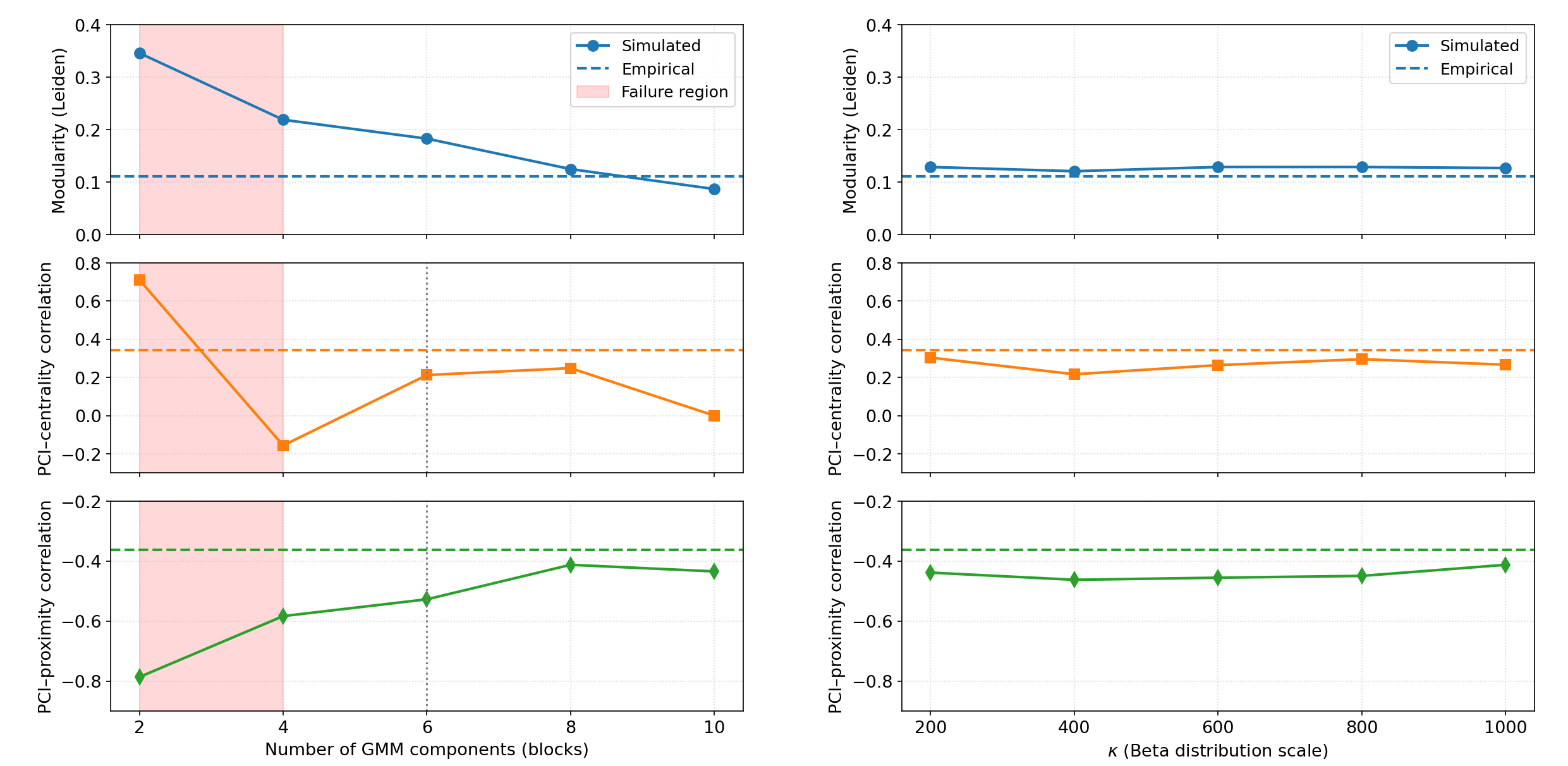}
        \caption{Sensitivity analysis for model simulation of the Product Space, including varying number of GMM components (left) and scale parameter for the Beta distribution representing within-block capability heterogeneity (right). Beta distribution results stable; GMM results fail for $n < 6$.}
    \end{figure}
\end{center}
\newpage

\section{Additional results for Section 6: Modeling economic complexity}

\subsection{Methodology in detail}

We begin with the export vector $\vec{x}_c$ containing the proportion of total exports occupied by each product, such that 

\begin{equation}
    \vec{x}_c = \begin{bmatrix}
        \frac{x_{p_1}}{\sum_{p \in P} x_p} \\
        \frac{x_{p_2}}{\sum_{p \in P} x_p} \\
        \vdots \\
        \frac{x_{p_{N_p}}}{\sum_{p \in P} x_p}
    \end{bmatrix}
\end{equation}

where $x_{p_i}$ is the total export volume of country $c$ in product $i$, $P$ is the set of all products, and $N_p = |P|$ is the total number of products. Given a set of capabilities $c^*$ possessed by country $c$, the model predicts the following form $\vec{x}^*_c$ for the export vector $\vec{x}_c$:
\begin{equation}
    \vec{x}^*_c = \begin{bmatrix}
        R(c^*, p_1) \\
         R(c^*, p_2) \\
         \vdots \\
          R(c^*, p_{N_p})
    \end{bmatrix}
\end{equation}
where $R(c^*, p_i)$ assumes the functional form given in Section 3. Thus define the set of capabilities $c^*$ possessed by country $c$ as the subset of the capability set $A$ which minimizes the Kullback-Leibler divergence between $\vec{x}_c$ and $\vec{x}^*_c$, denoted $KL(\vec{x}_c, \vec{x}^*_c)$:

\begin{equation}
    KL(\vec{x}_c || \vec{x}^*_c) = \sum_{i=1}^{N_p} (\vec{x}_c)_i \ln \frac{(\vec{x}_c)_i}{(\vec{x}^*_c)_i}.
\end{equation}

The Kullback-Leibler (KL) divergence is used to measure deviation between $\vec{x}_c$ (the true distribution) and $\vec{x}^*_c$ (the observed distribution) because each of $\vec{x}$ and $\vec{x}^*_c$ can be considered discrete probability distributions for choosing a certain product at random from the export basket of $c$, weighted by export volume. As such, the KL divergence provides a more detailed picture of distribution divergence than measures like Kolmogorov-Smirnov distance, which measure maximum distance between two cumulative distributions instead of divergence between every single point in the distribution; it also has the information-theoretic property of being equivalent to the negative log-likelihood of observing $\vec{x}_c$ given a theoretical distribution of $\vec{x}^*_c$, meaning that minimizing the K-L divergence is equivalent to maximizing the likelihood of $\vec{x}^*_c$ being the underlying set of capabilities. As our goal is to find a set of capabilities that most plausibly explains the empirically observed export basket, this is a very desirable property. \\

It is important to note that the products $p_1, p_2, ..., p_{N_p}$ appearing above are \it not \normalfont real-world products; instead, they are products generated by the model and sorted by $K_{p_i, 0}$, or the number of capabilities used to produce them, in ascending order. Each product $K_{p_i, 0}$ is assigned a product complexity equal to $K_{p_i,0}$ scaled to the range of the Product Complexity Index:

\begin{equation}
    K_{p_i, 0}' = \frac{K_{p_i, 0} - K_{min}}{K_{max} - K_{min}}\times (PCI_{max} - PCI_{min})
\end{equation}
where $K_{max}$ and $K_{min}$ are the maximum and minimum values of $K_{p_i, 0}$ respectively, and $PCI_{max}$ and $PCI_{min}$ are the maximum and minimum values of empirically-calculated PCI. This leads to a range of products whose complexities are distributed identically to the best-fitting GMM to the PCI data (see Section 3). Moreover, due to all products now being clearly ordered on a scale, we can interpolate the entries of the observed country export vector $\vec{x}_c$ by fitting a kernel density estimator (KDE) over the empirical export versus product complexity data:
\begin{equation}
    F(PCI) = \frac{1}{Nh}\sum_{i=1}^{N}K(\frac{PCI - PCI_i}{h})
\end{equation}
where $N$ is the number of empirically observed products (5008), $K$ is taken to be the Gaussian kernel
\begin{equation}
    K(x) = \frac{1}{\sqrt{2\pi}} e^{-x^2/2} 
\end{equation}
which provides an continuous estimation for the distribution of product exports versus PCI for any country, and $h$ is a \it bandwidth \normalfont parameter representing the standard deviation of each Gaussian kernel, estimated via the Improved Sheather-Jones algorithm (Jones, Marron and Sheather 1996), which finds via recursion a value of $h$ that minimizes

\begin{equation}
MISE(h) = E(\int (F_h(x) - F_{actual}(x))^2\ dx)  
\end{equation}
where $F_h(x)$ is the value predicted by the KDE for the export share of a product with $PCI = x$, $F_{actual}(x)$ is the product's actual export share, and the quantity being minimized is the mean integrated square error (MISE). \\

Thus, we take $\vec{x}_c$, the vector of empirically observed product exports, to be 
\begin{equation}
    \vec{x}_c = \begin{bmatrix}
        F(K'_{p_1, 0}) \\
        F(K'_{p_2, 0}) \\
        \vdots \\
        F(K'_{p_{N_p}, 0})
    \end{bmatrix}.
\end{equation}
We make the choice of using a KDE to interpolate theoretical product export shares rather than directly use empirically observed product export shares (e.g. by converting the PCIs of the 5008 products to a scale of capabilities, then generating capability sets) for several reasons. \\

First, the export vector for any country in any year is inherently noisy and idiosyncratic; indeed, less developed countries often do not export or export very little of a majority of products as shown by the right-skewed distribution of country diversity (Hausmann and Hidalgo 2011), leading to a sparse vector with many zeroes. Not only does this lead to a very rugged optimization landscape, it also leads to a complete breakdown of the KL divergence calculation, which requires nonzero probabilities. \\

Second, as demonstrated in Section 2, the mathematical interpretation of PCI is as an approximately optimal spectral clustering of a product-product similarity matrix akin to the Product Space; Section 5 confirms the intuition that proximity between two products in the Product Space may be explained by their requiring related capabilities, meaning that products with similar PCIs are more likely to be placed in close proximity in the Product Space, and thus more likely to be similar in terms of their capability makeup. This supports the interpretation of products with similar PCIs requiring similar capabilities, and thus the fact that products with similar PCIs will be exported at similar volumes, varying smoothly across the product complexity axis. A KDE which minimizes the expected mean-squared error via the Sheather-Jones algorithm creates the most accurate interpolation possible: if a country has observed export shares $x_a$ and $x_b$ for two products with PCIs $a$ and $b$, it tells us what volume of a product with PCI $c = \frac{a+b}{2}$ may be exported at. 

\subsection{Optimizing capabilities}
In the main body of the paper, we briefly describe a simulated annealing algorithm which optimizes for a set of capabilities which minimizes KL divergence between a country's predicted and actual export distributions; it does so through considering a binary vector of capabilities (of length 200, the total number of capabilities in the model: 8 blocks, with 25 capabilities per block), where the $i$th entry is $0$ if the country does not possess capability $i$ and $1$ otherwise, and flipping the entries of this vector one at a time. The algorithm is started with temperature $1$ which exponentially decays with cooling rate $0.95$, and repeated 5 times with the set of capabilities minimizing the KL divergence chosen at the end. Full code for the algorithm can be found in the GitHub repository provided in the main body of the paper, in the subsection "Code and data availability".

\subsection{Additional results for Section 6.1: From capabilities to economic development}

In the following subsection we present additional results based on simulations of the model for data in 2000, 2010 and 2015 in regards to 1) the goodness-of-fit of the model to empirical export distributions of countries, in terms of clarity (percentage reduction in KL divergence between predicted export vector and empirical export vector when compared with uniform distribution) and mean KL divergence; and 2) regressions between $\rho$, $\nu$, and ECI, with model specification identical to those in the main body of the paper. 
\newpage
\begin{table}[h]
\centering
\caption{Goodness-of-fit statistics for model on empirical export distributions, 2000-2015}
\begin{tabular}{lccc}
 \textbf{Year} & \textbf{Mean clarity}  & \textbf{Mean KL Divergence} \\
\midrule
\textbf{2000} & 0.205 & 1.767 \\
\textbf{2010} & 0.220 & 1.794 \\
\textbf{2015} & 0.150 & 1.532
\end{tabular}
\end{table}

\begin{table}[h]
\centering
\caption{Logit regression for $\rho$ and $\nu$ against ECI, 2000}
\begin{tabular}{lcccc}
\toprule
 & (1) & (2) & (3) & (4) \\
Variables & $\rho$ (ECI) & $\nu$ (ECI) & $\rho$ (No ECI) & $\nu$ (No ECI)\\
\midrule
Log GDP per capita & 0.126 & -0.188 & 0.300** & -0.375**  \\
 & (0.150) & (0.138) & (0.114) & (0.108) \\
Square log GDP per capita & -0.025 & -0.066* & -0.018 & -0.107* \\
& (0.073) & (0.067)  & (0.066) & (0.062)\\
ECI & 0.442*** & -0.423*** & & \\
& (0.217) & (0.197) & & \\
ECI squared & -0.1324 & -0.052 & & \\
& (0.130) & (0.122) & & \\
Population & -0.000 & -0.003** & 0.000 & -0.003** \\
& (0.001) & (0.001) & (0.001) & (0.001) \\
Investment-GDP ratio & 0.007 & -0.021 & 0.003 & -0.019 \\
& (0.018) & (0.015) & (0.018) & (0.015) \\
Export-GDP ratio &-0.005 & 0.012** & 0.004 & 0.013**  \\
& (0.006) & (0.005) & (0.006) & (0.005) \\   
\midrule
Observations & 165 & 165 & 165 & 165 \\
(Psd.) R-squared & 0.050 & 0.066 &0.033 & 0.057 \\
$\chi^2$-statistic, log-likelihood & 16.984** & 31.875*** & 11.308** & 27.153***\\
AIC & 347.0 & 470.1 & 348.6 & 470.9 \\
\bottomrule
\end{tabular}
\begin{tablenotes}
\small
\item Notes: The dependent variable is different categories of $\rho$ and $\nu$.  Robust standard errors (HC1) are in parentheses. *** p$\textless$0.01, ** p$\textless$0.05, * p$\textless$0.1.
\end{tablenotes}
\end{table}
\newpage
\begin{table}[h]
\centering
\caption{Logit regression for $\rho$ and $\nu$ against ECI, 2010}

\begin{tabular}{lcccc}
\toprule
 & (1) & (2) & (3) & (4) \\
Variables & $\rho$ (ECI) & $\nu$ (ECI) & $\rho$ (No ECI) & $\nu$ (No ECI)\\
\midrule
Log GDP per capita & -0.147 & 0.119 & 0.207* & -0.250**  \\
 & (0.170) & (0.162) & (0.121) & (0.115) \\
Square log GDP per capita & -0.084 & 0.003 & -0.013 & -0.061* \\
& (0.079) & (0.071)  & (0.071) & (0.067)\\
ECI & 0.674*** & -0.664*** & & \\
& (0.217) & (0.209) & & \\
ECI squared & 0.154 & -0.089 & & \\
& (0.124) & (0.114) & & \\
Population & -0.000 & -0.000** & 0.001 & -0.001** \\
& (0.001) & (0.001) & (0.001) & (0.001) \\
Investment-GDP ratio & 0.005 & 0.015 & -0.005 & 0.020 \\
& (0.018) & (0.016) & (0.018) & (0.015) \\
Export-GDP ratio &-0.007 & 0.004 & -0.007 & 0.006 \\
& (0.006) & (0.006) & (0.006) & (0.006) \\   
\midrule
Observations & 168 & 168 & 168 & 168 \\
(Psd.) R-squared & 0.035 & 0.040 & 0.009 & 0.018 \\
$\chi^2$-statistic, log-likelihood & 12.898* & 19.306*** & 3.358 & 8.794 \\
AIC & 381.1 & 483.6 & 386.7 & 490.1 \\
\bottomrule
\end{tabular}
\begin{tablenotes}
\small
\item Notes: The dependent variable is different categories of $\rho$ and $\nu$.  Robust standard errors (HC1) are in parentheses. *** p$\textless$0.01, ** p$\textless$0.05, * p$\textless$0.1.
\end{tablenotes}
\end{table}
\newpage
\begin{table}[h]
\centering
\caption{Logit regression for $\rho$ and $\nu$ against ECI, 2015}
\begin{tabular}{lcccc}
\toprule
 & (1) & (2) & (3) & (4) \\
Variables & $\rho$ (ECI) & $\nu$ (ECI) & $\rho$ (No ECI) & $\nu$ (No ECI)\\
\midrule
Log GDP per capita & 0.122 & 0.089 & 0.298** & -0.167  \\
 & (0.166) & (0.155) & (0.125) & (0.111) \\
Square log GDP per capita & 0.087 & 0.105 & 0.132* & 0.121* \\
& (0.078) & (0.070)  & (0.075) & (0.066)\\
ECI & 0.374* & -0.545** & & \\
& (0.240) & (0.230) & & \\
ECI squared & 0.220 & 0.205 & & \\
& (0.146) & (0.130) & & \\
Population & -0.001 & -0.002** & -0.001 & -0.003** \\
& (0.001) & (0.001) & (0.001) & (0.001) \\
Investment-GDP ratio & 0.007 & 0.008 & 0.004 & 0.014 \\
& (0.017) & (0.016) & (0.017) & (0.016) \\
Export-GDP ratio & 0.002 & -0.003 & 0.004 & -0.004 \\
& (0.007) & (0.006) & (0.006) & (0.005) \\   
\midrule
Observations & 162 & 162 & 162 & 162 \\
(Psd.) R-squared & 0.048 & 0.049 & 0.037 & 0.031 \\
$\chi^2$-statistic, log-likelihood & 17.405** & 22.427*** & 13.279** & 14.061** \\
AIC & 363.7 & 456.6 & 363.9 & 460.9 \\
\bottomrule
\end{tabular}
\begin{tablenotes}
\small
\item Notes: The dependent variable is different categories of $\rho$ and $\nu$.  Robust standard errors (HC1) are in parentheses. *** p$\textless$0.01, ** p$\textless$0.05, * p$\textless$0.1.
\end{tablenotes}
\end{table}

\subsection{Additional results for Section 6.2: From capabilities to complexity}

This subsection will present additional results which shed light on 1) the connection between the measure of economic complexity derived from our model of capabilities and the ECI, through a quantitative table of correlation coefficients throughout different years and through a qualitative scatterplot between the two statistics, specifically highlighting countries of interest and how they compare in the two metrics; and 2) how well our measure of complexity performs in forecasting growth over different time periods and with different start years compared to the ECI, and under different regression specifications. \\

Only the correlation coefficients between ECI and the two measures of complexity of our model (number of capabilities, average capability complexity) are included in the following table. \newpage
\begin{table}[h]
\caption{Correlation between different complexity measures, 2000-2015}
\begin{tabular}{lccc}
 \textbf{Year} & \textbf{ECI and avg. cap.}  & \textbf{ECI and num. cap.} & \textbf{avg. cap. and num. cap.} \\
\midrule
$2000$ & 0.588 & 0.364 & 0.417  \\ 
$2010$ & 0.537 & 0.298 & 0.358 \\ 
$2015$ & 0.654 & 0.288 & 0.308  \\ 
\end{tabular}
\end{table}

\begin{center}
    \begin{figure}[h]
        \includegraphics[width=13cm]{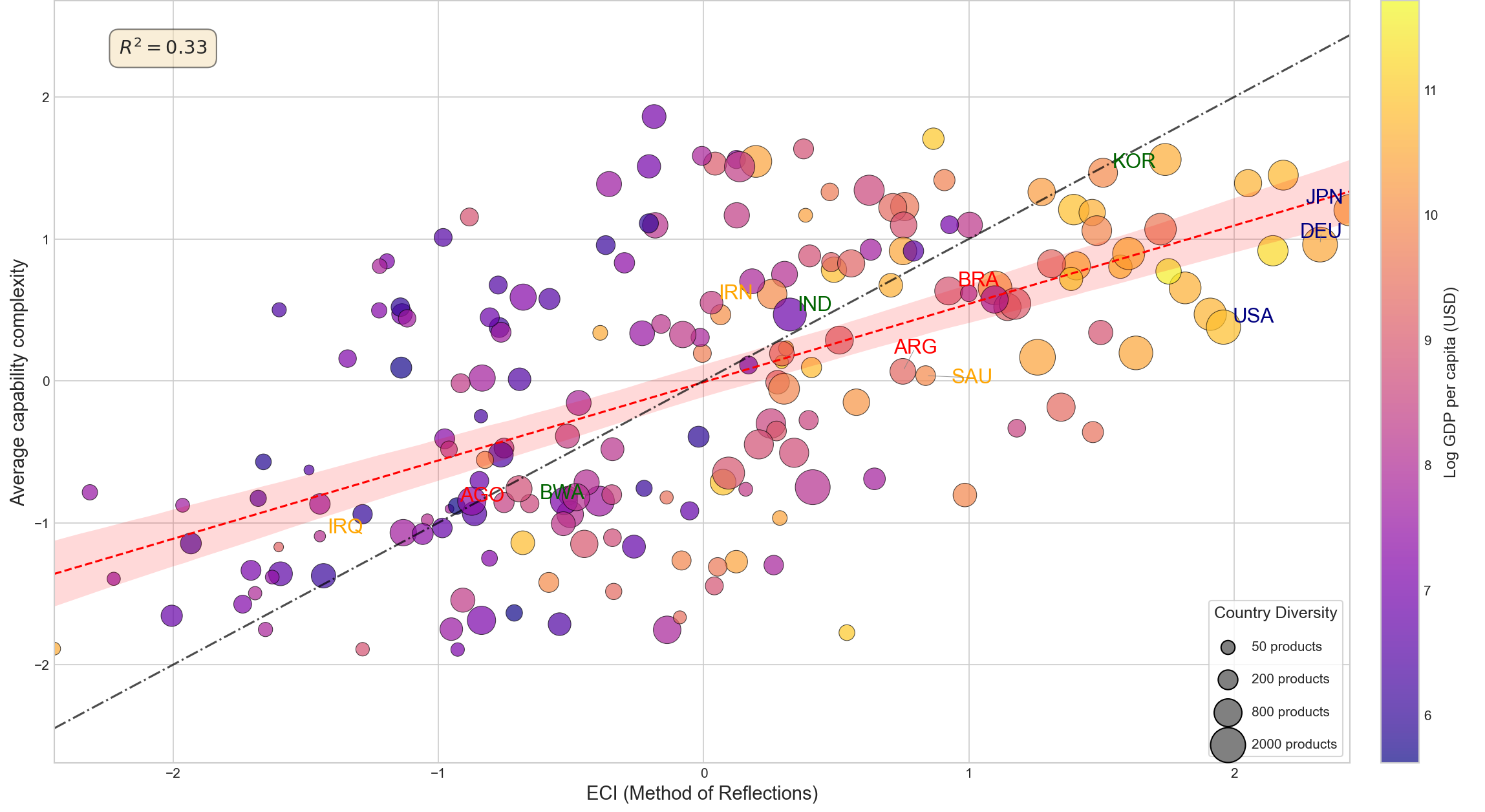}
        \caption{Scatter plot of ECI (x-axis) against average capability complexity (y-axis), both normalized. $R = 0.56; R^2 = 0.33$. Size of data point is country diversity; color is log GDP per capita, which both variables are correlated with. Black dotted line is $x=y$; red dotted line is line of best fit}
    \end{figure}
\end{center}

In particular, specific countries which may prove informative in this scatter-plot are labelled directly: highly-developed economies in dark blue (Japan (JPN), Germany (DEU), and the US (USA), being highly complex according to both measures), rapidly-growing economies in green (Korea (KOR), India (IND), and Botswana (BWA)), stagnating economies in red (Angola (AGO), Argentina (ARG), and Brazil (BRA)), and resource-rich, highly-specialized economies in yellow (Iraq (IRQ), Iran (IRN), and Saudi Arabia (SAU)). We observe that the model predicts a lower complexity for highly-developed economies than the ECI, except for Korea, which grew the quickest out of all four from 2005 to 2025; the other economies are roughly scattered around the line $x=y$.  \\

In the main body of the paper, we mentioned four regressions with base year 2005 with dependent variable being 20-year average growth (2005-2025), under the following specifications:

\begin{enumerate}
    \item Original ECI and log GDP per capita, as in Hidalgo and Hausmann's original paper.
    \item Original ECI and all explanatory variables above \textbf{except} diversity and population (led to multicollinearity).
    \item Average capability complexity and log GDP per capita.
    \item Average capability complexity and all explanatory variables above \textbf{except} diversity (led to multicollinearity).
\end{enumerate}

We presented regressions 2 and 4 as a justification of the ability of average capability complexity to forecast growth even when accounting for the most rigorous set of controls. For completeness, we now present regressions 1 and 3. 

\begin{table}[h]
\centering
\caption{Simple growth regressions (2005 starting year, 20‑year horizon)}
\begin{tabular}{l c c}
\toprule
 & (1) ECI & (2) Avg. capability complexity \\
\midrule
Growth, 20-year average & 0.388* & 0.304** \\
 & (0.203) & (0.143) \\
Log GDP per capita & -0.568*** & -0.444*** \\
 & (0.137) & (0.088) \\
Constant & 6.667*** & 5.605*** \\
 & (1.185) & (0.799) \\
\midrule
Observations & 165 & 165 \\
R-squared & 0.104 & 0.103 \\
\bottomrule
\end{tabular}
\medskip
{\small \textit{Notes:} Dependent variable is average annual GDP per capita growth 2006–2023. Robust standard errors in parentheses. *** p<0.01, ** p<0.05, * p<0.1.}
\end{table}

The following present regression results for average capability complexity and ECI with the same set of explanatory variables as in regression 4 of the main body of the paper (log GDP per capita, investment-GDP ratio, export-GDP ratio, population), over different time-periods if data allows (5-, 10- and 20-year average growth rates). 

\begin{table}[h]
\centering
\caption{ECI regressions with full controls (investment, export, log GDP per capita)}
\begin{tabular}{l c c c c c c c}
\toprule
Year & Horizon & ECI & Log GDP pc & Investment-GDP & Export-GDP & Obs. & $R^2$ \\
\midrule
2000 & 5-yr & 0.486 & -0.718** & 0.100** & 0.005 & 165 & 0.089 \\
 & & (0.454) & (0.282) & (0.049) & (0.014) & & \\
2000 & 10-yr & 0.402 & -0.871*** & 0.069** & 0.003 & 165 & 0.173 \\
 & & (0.302) & (0.211) & (0.029) & (0.010) & & \\
2000 & 20-yr & 0.360** & -0.733*** & 0.038* & 0.002 & 165 & 0.221 \\
 & & (0.166) & (0.127) & (0.015) & (0.006) & & \\
2005 & 5-yr & 0.301 & -0.944*** & 0.070** & 0.005 & 165 & 0.184 \\
 & & (0.250) & (0.201) & (0.030) & (0.009) & & \\
2005 & 10-yr & 0.251 & -0.829*** & 0.045** & 0.007 & 165 & 0.208 \\
 & & (0.200) & (0.151) & (0.022) & (0.007) & & \\
2010 & 5-yr & -0.041 & -0.541** & 0.035 & 0.005 & 168 & 0.102 \\
 & & (0.280) & (0.221) & (0.024) & (0.008) & & \\
2010 & 10-yr & 0.369 & -0.625** & 0.029 & -0.002 & 168 & 0.109 \\
 & & (0.270) & (0.184) & (0.021) & (0.006) & & \\
2015 & 5-yr & 0.988** & -0.896*** & -0.002 & 0.002 & 162 & 0.083 \\
 & & (0.329) & (0.211) & (0.026) & (0.006) & & \\
\bottomrule
\end{tabular}
\medskip
{\small \textit{Notes:} Dependent variable is average annual GDP per capita growth over the indicated horizon. 20‑year results for 2005 are in the main text (Table 7). Robust standard errors in parentheses. *** p\textless0.01, ** p\textless0.05, * p\textless0.1.}
\end{table}

\begin{table}[h]
\centering
\caption{ACC regressions with full controls (investment, export, log GDP per capita, population)}
\begin{tabular}{l c c c c c c c c}
\toprule
Year & Horizon & ACC & Log GDP pc & Investment-GDP & Export-GDP & Population & Obs. & $R^2$ \\
\midrule
2000 & 5-yr & 0.384 & -0.594*** & 0.093* & 0.007 & 0.003** & 165 & 0.095 \\
 & & (0.317) & (0.227) & (0.050) & (0.014) & (0.001) & & \\
2000 & 10-yr & 0.386 & -0.782*** & 0.062** & 0.006 & 0.004** & 165 & 0.204 \\
 & & (0.237) & (0.165) & (0.028) & (0.010) & (0.001) & & \\
2000 & 20-yr & 0.401** & -0.671*** & 0.031** & 0.005 & 0.003** & 165 & 0.291 \\
 & & (0.128) & (0.099) & (0.014) & (0.006) & (0.001) & & \\
2005 & 5-yr & 0.148 & -0.810*** & 0.059** & 0.007 & 0.004** & 165 & 0.208 \\
 & & (0.199) & (0.146) & (0.030) & (0.009) & (0.002) & & \\
2005 & 10-yr & 0.169 & -0.723*** & 0.034 & 0.009 & 0.004** & 165 & 0.250 \\
 & & (0.155) & (0.105) & (0.021) & (0.007) & (0.001) & & \\
2010 & 5-yr & 0.041 & -0.574*** & 0.035** & 0.005 & --- & 168 & 0.102 \\
 & & (0.223) & (0.130) & (0.025) & (0.008) & & & \\
2010 & 10-yr & 0.269 & -0.495*** & 0.027 & -0.004 & --- & 168 & 0.105 \\
 & & (0.194) & (0.107) & (0.022) & (0.006) & & & \\
2015 & 5-yr & 0.290 & -0.501*** & -0.005 & 0.004 & --- & 162 & 0.024 \\
 & & (0.315) & (0.176) & (0.028) & (0.007) & & & \\
\bottomrule
\end{tabular}
\medskip
{\small \textit{Notes:} "ACC" refers to average capability complexity. Dependent variable is average annual GDP per capita growth over the indicated horizon. 20‑year results for 2005 are in the main text (Table 8). Population was not included in the 2010 and 2015 regressions due to multicollinearity (original tables have no population row). Robust standard errors in parentheses. *** p\textless0.01, ** p\textless0.05, * p\textless0.1.}
\end{table}

\end{appendices}

\end{document}